\documentclass[useAMS,usenatbib]{mn2e}

\usepackage{graphicx}
\usepackage{epsfig}
\usepackage{amssymb,amsmath}
\usepackage{multirow}
\usepackage{mathrsfs}
\usepackage{times}
\usepackage{setspace}
\usepackage{color}

\newcommand{\black}[1]{{\textcolor{black}{#1}}}

\usepackage[colorlinks, breaklinks, citecolor=blue]{hyperref}

\newcommand{\noopsort}[1]{}
\newcommand{\emcee}{\texttt{emcee}}
\newcommand{\cs}{\texttt{cluster\_slug}}
\newcommand{\slug}{\texttt{slug}}
\newcommand{\vectheta}{\boldsymbol{\theta}}
\newcommand{\vecL}{\boldsymbol{L}}

\newcommand{\vech}{\boldsymbol{h}}
\newcommand{\vecsigma}{\boldsymbol{\sigma}}
\newcommand{\nobs}{N_{\mathrm{obs}}}
\newcommand{\gmdd}{{\gamma_{\rm mdd}}}
\newcommand{\tmdd}{{T_{\rm mdd,min}}}

\title[SLUG IV: Forward-Modelling Star Cluster Demographics]{SLUG IV: A Novel Forward-Modelling Method to Derive the Demographics of Star Clusters}

\author[Krumholz et al.]{Mark R. Krumholz$^1$\thanks{mark.krumholz@anu.edu.au},
Angela Adamo$^2$, Michele Fumagalli$^3$, Daniela Calzetti$^4$
\\ \\
$^1$ Research School of Astronomy \& Astrophysics, Australian National University, Canberra, ACT 2612, Australia\\
$^2$ Department of Astronomy, Oskar Klein Centre, Stockholm University, SE-10691 Stockholm, Sweden\\
$^3$ Institute for Computational Cosmology and Centre for Extragalactic Astronomy, Department of Physics, Durham University, South Road, Durham, \\
DH1 3LE, UK\\
$^4$ Department of Astronomy, University of Massachusetts -- Amherst, Amherst, MA, USA
}

\begin{document}
\maketitle
\label{firstpage}
\begin{abstract}
We describe a novel method for determining the demographics of a population of star clusters, for example distributions of cluster mass and age, from unresolved photometry. This method has a number of desirable properties: it fully exploits all the information available in a data set without any binning, correctly accounts for both measurement error and sample incompleteness, naturally handles heterogenous data (for example fields that have been imaged with different sets of filters or to different depths), marginalises over uncertain extinctions, and returns the full posterior distributions of the parameters describing star cluster demographics. We demonstrate the method using mock star cluster catalogs and show that our method is robust and accurate, and that it can recover the demographics of star cluster populations significantly better than traditional fitting methods. For realistic sample sizes, our method is sufficiently powerful that its accuracy is ultimately limited by the accuracy of the underlying physical models for stellar evolution and interstellar dust, rather than by statistical uncertainties. Our method is implemented as part of the Stochastically Lighting Up Galaxies (\texttt{slug}) stellar populations code, and is freely available.
%\vspace{0.5in}
\end{abstract}

\begin{keywords}
methods: data analysis --- methods: statistical --- galaxies: star clusters: general --- techniques: photometric
\end{keywords}

%\clearpage

\section{Introduction}
\label{sec:intro}

Stars form in regions where the stellar density is vastly higher than the mean for the galactic field. Over tens to hundreds of Myr after their formation, stars disperse from these birthplaces, leaving behind a small fraction of long-lived, gravitationally-bound old star clusters. This process of formation and dispersal encodes a great deal of physics regarding the formation of stars, the expulsion of gas from star-forming clouds, and the dynamical evolution of stellar systems in a galactic potential. For recent reviews, see \citet{krumholz14c}, \citet{krumholz14e}, and \citet{longmore14a}.

Because of the physics it encodes, the distribution of star cluster ages and masses has long been an important topic of study, both observationally and theoretically. Theoretical models for cluster dispersal have emphasised processes such as gas expulsion \citep[e.g.,][]{baumgardt07a, parmentier08a, krumholz09d, fall10a, murray10a}, tidal disruption of clusters after gas expulsion \citep[e.g.,][]{lamers05a, gieles07a, kruijssen09a, kruijssen12a, kruijssen12b, elmegreen10b}, and two-body relaxation and evaporation over long timescales \citep[e.g.,][]{fall01a}. These models predict a variety of functional forms for the joint age and mass distribution of surviving star clusters. Observational studies have attempted to measure these quantities for star clusters in the Milky Way \citep{williams97a, lada03a, borissova11a}, the Magellanic Clouds \citep{hunter03a, rafelski05a, chandar10a, popescu12a}, and more distant systems \citep{zhang99a, larsen02a, goddard10a, chandar10b, chandar11a, bastian12a, fall12a, fouesneau12a, fouesneau14a, de-meulenaer15a, krumholz15c, johnson16a, johnson17b, adamo17a, messa18a}, with the goal of testing the predictions of these models.

Because it is not at present possible to resolve the individual stars in young star clusters beyond the Milky Way and its few nearest neighbours, observational studies that go beyond samples of a few galaxies are generally restricted to working with unresolved light, where the raw data consist of measurements of luminosities in some set of filters for each star cluster. Consequently, there is an urgent need for robust statistical techniques to derive the physical properties of star cluster populations from such integrated light data; the method we introduce below is intended for this type of analysis.

The traditional approach for analysing these data is to assign an age and mass to each cluster by comparing their unresolved luminosities and colours to a set of evolutionary tracks for simple stellar populations, with the best-fitting mass and age determined by $\chi^2$ minimisation or a similar procedure. Once the masses and ages are determined, the clusters are placed in mass and age bins, and the distribution in the population as a whole can, in principle, be measured. However, such an approach encounters several difficulties. First, the process of binning inevitably discards some of the information present in the original data, and fitting parameters to binned distributions can introduce severe biases \citep[e.g.,][]{maschberger09a}. Second, at low masses, and for certain age and colour combinations even at higher masses, the assignment of mass and age to an individual cluster is highly uncertain, and the errors in the assignments are not well-approximated by simple Gaussians. Instead, the posterior probability distribution function (PDF) of mass and age can have a complex, multi-peaked shape \citep{popescu09a, popescu10a, popescu10b, fouesneau14a, de-meulenaer15a, krumholz15c}. A single best fit mass and age may be a very poor representation of the PDF for a single cluster, but the process of assigning a cluster to a single bin ignores this complexity. Third and most seriously, determining the properties of the population as a whole requires considering the completeness of the observed sample. Variations in whether and how one takes completeness into account can lead to quite different inferences in the final physical distributions \citep[e.g.,][]{lamers09a}. Part of the reason for this sensitivity is that completeness is a function of the luminosity and surface brightness profile of the cluster, the background, and the level of crowding in the image, leading to a completeness that has a complex functional form in mass-age-extinction space.

The simplest approach to handling the problem of completeness is to be extremely conservative, and discard all data in regions of parameter space where the observations are not complete or nearly so. However, this invariably requires one to discard much of the available data. A somewhat more sophisticated approach is forward modelling: rather than deriving the mass and age distribution of the population from estimates of mass age for individual clusters, one could instead consider a proposed distribution of masses and ages, predict the resulting photometry distribution including the effects of incompleteness, and adjust parameters of the mass and age distribution until they match the observations. Approaches of this type are widely used in astronomy, for example to infer star formation histories or stellar mass distributions from observed colour-magnitude diagrams (CMDs; e.g., \citealt{dolphin02a, harris09a, weisz13a, conroy16a}; see \citealt{cervino13a} for a review). However, methods of this type have not previously been applied to deriving the properties of populations of star clusters, at least in part due a unique challenge not present in other applications. In existing applications such as CMD fitting, the forward model is deterministic, i.e., for a given stellar mass, age, and other properties, there is a single predicted colour and magnitude. This is not the case for star clusters with masses $\lesssim 3000$ $M_\odot$, because such clusters are too small to fully sample the stellar initial mass function \citep[e.g.][]{cervino04a, cervino06a, da-silva12a}. As a result, two clusters of the same total mass and age can produce wildly different luminosities and colours. This means that the forward model is not deterministic, but instead depends on an additional random variable that couples non-linearly with the deterministic variables like cluster mass and age. This situation presents computational challenges that are not addressed by existing methods.

In this paper we introduce a new approach for determining the distribution of the properties of star clusters from unresolved photometry that allows us to consider arbitrary functional forms for distributions of mass, age, and extinction, and to exploit all the information available in heterogenous data (i.e., data where not all fields are observed with the same filters or to the same depth). Crucially, it naturally accounts for both incomplete observations and the uncertainties in the assignment of masses and ages to individual clusters that arise when the mapping between physical properties and luminosity is non-deterministic due to finite sampling. The essence of our approach is to consider a proposed distribution of physical parameters, determine the corresponding luminosity distribution in a probabilistic way so that we preserve the non-unique mapping between physical properties and photometry, apply the completeness function in observed luminosity space, and then compare to the data. We then adjust the underlying physical distribution until the best match to the observations is found. We implement this method using fast numerical algorithms that enable us to identify the parameters describing a cluster distribution on a workstation-level computer in $\sim 10$ hours of computing time. The software is based on the Stochastically Lighting Up Galaxies (\slug) software suite \citep{da-silva12a, da-silva14b, krumholz15b}, and is freely available from the \slug~website, \url{http://www.slugsps.com/cluster-population-pipeline}.

The plan for the remainder of this paper is as follows. In \autoref{sec:method}, we describe our new method and the computational techniques we use to implement it. In \autoref{sec:mock} we test the method on mock data to verify its accuracy and demonstrate its capabilities, and in \autoref{sec:conventional} we compare the performance of our new method to more conventional approaches. We summarise our findings in \autoref{sec:conclusions}.

\section{Method}
\label{sec:method}

\subsection{Statement of the Problem}

Our goal is to infer the mass and age distribution of a population of star clusters in a galaxy for which we have a sample of star clusters observed in a some set of photometric bands; we must include extinction as an additional nuisance parameter, which we will marginalise over. The method we develop to achieve this goal generalises and extends the one proposed by \citet{weisz13a} for inferring the initial mass function of a resolved population of individual stars. Formally, let $g(M,T,A_V\mid\vectheta)$ be the joint distribution of mass, age, and visual extinction for the underlying population, which depends on a vector of parameters $\vectheta$.\footnote{Metallicity is another potential physical parameter, but for simplicity we will assume that the cluster-to-cluster variation in metallicity is small enough that its effects can be neglected. This assumption is particularly likely to be valid for optical data, since metallicity has relatively little effect on optical bands, and mostly affect near-infrared colours \citep{anders04a}. The generalisation to include metallicity as a parameter is straightforward.} For example, if we were to assume that the mass, age, and $A_V$ distributions are separable powerlaws, then $\vectheta$ would contain the minimum, maximum, and slope of each powerlaw. We wish to infer a posterior distribution for $\vectheta$. Since the true size of the cluster population is not known \textit{a priori}, and our observations are inevitably incomplete at the low luminosity end of the distribution, we must also treat the number of clusters present $N_c$ as a parameter of the model, although we will see below that it is more convenient to transform to a different variable.

The data from which we will make this inference consists of a set of $\nobs$ observed star clusters, and for the $i$th star cluster we observe its absolute magnitude or luminosity $L_{F,i}$ in $N_F$ different photometric filters $F$, measured with some photometric error $\sigma_{F,i}$, which we take to be known and Gaussian-distributed. For notational compactness, let $\vecL_i$ and $\vecsigma_i$ be the luminosities / absolute magnitudes and the corresponding errors for the $i$th cluster in all $N_F$ filters, and $\{\vecL_i\}$ and $\{\vecsigma_i\}$ be the set of all such observed luminosities and uncertainties for all clusters in every filter. 

Finally, let us assume that each of our observations has a known completeness function described by $P_{\mathrm{obs}, i}(\vecL')$. This function is the probability that a cluster of intrinsic luminosity $\vecL'$, observed in the same manner as observed cluster $i$ (i.e., with the same integration time and set of filters, in a field at the same distance) will be included in the sample.\footnote{Here and throughout we use lowercase $p$ to denote probability distribution functions, and uppercase $P$ to denote simple, dimensionless probabilities.} Note that $\vecL'$ is distinct from the quantity $\vecL$ introduced in the previous paragraph: the former is the true luminosity of a cluster, while the latter is the measured luminosity, which is slightly different due to observational error. A simple magnitude limit corresponds to $P_{{\rm obs},i}(\vecL')$ being a step function. In practice this function must be determined by artificial cluster tests or the like. Note that we explicitly allow for the possibility that different sets of observations may have different completeness limits, for example if we are combining data from two different galaxies at different distances, or from two fields within the same galaxy that were observed to different depths.

\subsection{Posterior Probability for a Cluster Population}

As usual in a Bayesian approach, we write the posterior probability distribution of the model parameters $(\vectheta, N_c)$ given the data as the product of the prior probabilities with the likelihood function, i.e.,\footnote{A note on notation: since all probability distributions can be properly normalised by requiring that their integrals be unity, in what follows we usually omit normalisation constants and write out all dependencies as proportionalities. The only exceptions are cases where we retain the normalisation constant for clarity.}
\begin{eqnarray}
\lefteqn{p(\vectheta, N_c \mid \{\vecL_i\}, \{\vecsigma_i\}, \nobs) }\qquad
\nonumber \\
& \propto &\,  p(\{\vecL_i\}, \nobs \mid \vectheta, N_c, \{\vecsigma_i\}) \, p_{\mathrm{prior}}(\vectheta, N_c).
\end{eqnarray}
We remind readers that, in this equation $\vectheta$ is the vector of parameters describing the distribution of star cluster properties, $N_c$ is the true number of clusters in the observed region, $\{\vecL_i\}$ and $\{\vecsigma_i\}$ are the vector of observed cluster luminosities or magnitudes and their corresponding uncertainties, and $\nobs$ is the number of observed clusters. The likelihood function $p(\{\vecL_i\}, \nobs \mid \vectheta, N_c, \{\vecsigma_i\})$ is simply the probability density of the data given the model and the observational errors, while $p_{\mathrm{prior}}(\vectheta, N_c)$ is the prior probability distribution for the parameters $\vectheta$ and $N_c$, and $p(\vectheta, N_c \mid \{\vecL_i\}, \{\vecsigma_i\}, \nobs)$ is the posterior PDF that we are attempting to compute. To evaluate it, we assume that the observed luminosity of each cluster represents an independent draw from an underlying distribution of star cluster luminosities, $p_L(\vecL \mid \vectheta, \vecsigma)$; note that $\vecL$ here is the observed luminosity, not the intrinsic one, and because the uncertainties $\vecsigma$ are not the same from one measurement to another, the luminosity distributions for each cluster are not identical. We defer a calculation of $p_L(\vecL \mid \vectheta, \vecsigma)$ to the next section. 

We assume that both the intrinsic luminosities of clusters and the observational errors on them are uncorrelated.\footnote{\black{The assumption of uncorrelated noise may not be strictly true in a real observation, since the dominant uncertainty in real observations is usually the aperture correction. This may lead to errors that depend on the level of crowding or background, and thus are correlated with respect to the locations of clusters within the target galaxy. However, this would represent a correlation of error with cluster position. As long as there is no correlation of the error with cluster physical properties, this does not matter for our purposes.}} Under this assumption, and since the number of observed clusters $\nobs$ is also an independent variable, we can write the likelihood function for the cluster population as a product of the probability distributions for individual clusters and for $\nobs$,
\begin{eqnarray}
\lefteqn{p(\{\vecL_i\}, \nobs \mid \vectheta, N_c, \{\vecsigma_i\}) } \qquad\nonumber \\
& \propto & \, P_N(\nobs \mid \vectheta, N_c) \prod_{i=1}^{\nobs} p_L(\vecL_i \mid \vectheta, \vecsigma_i).
\label{eq:likelihood1}
\end{eqnarray}
Here $P_N(\nobs\mid \vectheta, N_c)$ is the probability that we will observe $\nobs$ clusters from a population of $N_c$ whose intrinsic luminosity distribution is parameterised by $\vectheta$. 

To determine $P_N(\nobs \mid \vectheta, N_c)$, first consider the simplest case where the observed clusters all come from a single field imaged with a single set of filters and uniform sensitivity across it. In this case, there is a single completeness function $P_\mathrm{obs}(\vecL')$, and for a cluster population with a distribution of intrinsic luminosities $p_{L'}(\vecL' \mid \vectheta)$, for any set of population parameters $\vectheta$ there is a single probability
\begin{equation}
P_{\rm obs}(\vectheta) = \int P_\mathrm{obs} (\vecL') \, p_{L'}(\vecL' \mid \vectheta) \, d\vecL'
\end{equation}
that a randomly-selected cluster will be observed. In this case the number of clusters we expect to observe is $N_{\rm ex} = P_{\rm obs}(\vectheta) N_c$, and since each observation of one of the $N_c$ clusters present is an independent experiment, the actual number observed must be Poisson-distributed:
\begin{equation}
\label{eq:poisson}
P_N(\nobs \mid N_{\rm ex}) = \frac{N_{\mathrm{ex}}^{\nobs} e^{-N_\mathrm{ex}}}{\nobs!}.
\end{equation}
In this expression we have suppressed the dependence of $N_{\rm ex}$ on $\vectheta$ for the sake of compactness.

Now consider the more general case where we have multiple fields with different sensitivities and filter sets, and thus different completeness functions. For each such observation $j$ there will be some number of clusters $N_{\mathrm{ex},j}$ that one would expect to detect, which is a function of both the true number of clusters in the observed field and the observational completeness function for it. The number of clusters $N_{\mathrm{obs},j}$ that is actually observed in each field will then be Poisson-distributed per \autoref{eq:poisson}, and the total number of clusters expected in the full catalogue of all fields is just $N_{\rm ex} = \sum_j N_{{\rm ex},j}$. However, the sum of a number of random variables that are each drawn from a Poisson distribution is itself Poisson-distributed. Thus $P_N(\nobs \mid \vectheta, N_c)$ must be distributed following \autoref{eq:poisson} even for heterogenous observations.

Because $P_N(\nobs \mid \vectheta, N_c)$ depends only on $N_{\rm ex}$, it is convenient to eliminate $N_c$ in favour $N_{\rm ex}$ as the variable for which we will seek a posterior PDF.  That is, rather than trying to compute $p(\vectheta, N_c \mid \{\vecL_i\}, \{\vecsigma_i\}, \nobs)$, we will instead compute
\begin{eqnarray}
\lefteqn{p(\vectheta, N_{\rm ex} \mid \{\vecL_i\}, \{\vecsigma_i\}, \nobs) }\qquad
\nonumber \\
& \propto &\, P_N(\nobs \mid N_{\rm ex}) \prod_{i=1}^{\nobs} p_L(\vecL_i \mid \vectheta, \vecsigma_i).
\end{eqnarray}
We have therefore written the likelihood function for our observed cluster population in terms of the luminosity distribution for a single cluster and the expected number of observed clusters, with the dependence on $N_{\rm ex}$ separable from that on $\vectheta$.
Of course, once one has determined the posterior distributions of $\vectheta$ and $N_{\rm ex}$, one could use these to obtain the posterior distribution of $N_c$. In practice, however, this is unlikely to be interesting, for the following reason: star cluster mass functions are invariably observed to be steep, such that by number most clusters have low masses. Thus the value of $N_c$ will depend strongly on the shape of the mass function at low masses. Since real extragalactic observations invariably become incomplete at masses significantly larger than the smallest possible star cluster mass, any parameters we introduced to describe the shape of the lower part of the mass function (e.g., a lower mass cutoff), will not be constrained by the observations, and since $N_c$ depends critically upon them, it will be unconstrained by the observations as well.

\subsection{The Distribution of Observed Luminosities for Individual Clusters}

The final step in our derivation is to compute the distribution of observed luminosities for an individual cluster, $p_L(\vecL \mid \vectheta, \vecsigma)$, where we remind readers that $\vecL$ is the vector of \textit{observed} luminosities, which are the result of taking the true luminosity $\vecL'$ and measuring it with some finite error $\vecsigma$. To do so, we assume that there there is a distribution of \textit{intrinsic} cluster luminosities $\vecL'$ that is identical for every cluster, and that depends only the the model parameters: $p_{L'}(\vecL' \mid \vectheta)$. We can then obtain the observed luminosity distribution by marginalising over the intrinsic luminosity of each cluster:
\begin{eqnarray}
\lefteqn{p_L(\vecL \mid \vectheta, \vecsigma) } \quad
\nonumber \\
& \propto &
\,  \int p_{L'}(\vecL' \mid \vectheta) \, p(\vecL \mid \vecL',\vecsigma)  \, P_{\mathrm{obs}}(\vecL')
\, d\vecL',
\label{eq:pl1}
\end{eqnarray}
where $p(\vecL \mid \vecL', \sigma)$ is the probability that a cluster of intrinsic luminosity $\vecL'$ will yield an observed luminosity $\vecL$ when measured with uncertainty $\vecsigma$. Under our assumption that the observational uncertainties are Gaussian, this is
\begin{eqnarray}
p(\vecL \mid \vecL', \vecsigma) & = & \frac{\left(2\pi\right)^{-N_F/2}}{\prod_{n=1}^{N_F}\sigma_n} \exp\left[-\sum_{n=1}^{N_F}\frac{\left(L_n-L'_n\right)^2}{2\sigma_n^2}\right]
\nonumber \\
& \equiv &
\mathcal{N}(\vecL \mid \vecL', \vecsigma),
\label{eq:obs_uncertainty}
\end{eqnarray}
where the sum runs over all $N_F$ filters, and we have introduced the notation $\mathcal{N}(\mathbf{x} \mid \mathbf{x}_0, \vecsigma)$ to represent the usual multi-dimensional normal distribution centred on $\mathbf{x_0}$ with standard deviation $\vecsigma$ and no covariance, evaluated at position $\mathbf{x}$.

We estimate the intrinsic luminosity distribution convolved with the probability of being observed using the method described by \citet{krumholz15b}, and implemented in the \cs~module in the \slug~software package. Specifically, given a library of simulated clusters, where cluster $j$ has a mass $M_j$, age $T_j$, extinction $A_{V,j}$, and a vector of luminosities $\vecL'_j$, we write the intrinsic luminosity distribution using a kernel density estimation model,
\begin{equation}
\label{eq:kde}
p_{L'}(\vecL' \mid \vectheta) \, P_{\mathrm{obs}}(\vecL') \propto \sum_{j=1}^{N_{\rm lib}} w_j(\vectheta) \mathcal{N}(\vecL' \mid \vecL'_j, \vech).
\end{equation}
Here the sum runs over all $N_{\rm lib}$ clusters in the simulation library, $\vech$ is the bandwidth of the kernel density estimation, and the weights $w_j$ are given by
\begin{equation}
\label{eq:wgts}
w_j(\vectheta) = P_{\rm obs}(\vecL'_j) \frac{g(M_j, T_j, A_{V,j} \mid \vectheta) }{p_{\mathrm{lib}}(M_j, T_j, A_{V,j})},
\end{equation}
where $g(M, T, A_V \mid \vectheta)$ is the proposed distribution of mass, age, and extinction, and $p_{\mathrm{lib}}(M, T, A_V)$ is the distribution from which the library was sampled.

For the purposes of developing intuition, it is helpful to examine the weight factors $w_j(\vectheta)$ factors in a bit more detail. The meaning of the first term, $P_{\rm obs}(\vecL'_j)$, is simple: it simply down-weights the contribution of each library cluster to the observed luminosity distribution by the probability that will actually be observed. The factor $p_{\mathrm{lib}}(M, T, A_V)$ in the denominator of \autoref{eq:wgts} simply represents the frequency with which we drew a particular combination of $(M, T, A_V)$ in the process of constructing the library; that is, the number of sample library clusters that fall into a particular infinitesimal range in mass, age, and extinction is just proportional to $p_{\mathrm{lib}}(M, T, A_V)$. By contrast, $g(M, T, A_V \mid \vectheta)$ is the number of sampled points that we \textit{would} have had in that bin if our library had been drawn from the distribution described by the parameters $\vectheta$. Thus the ratio of these two terms, to which $w_j(\vectheta)$ is proportional, simply represents the ratio of the number of clusters we \textit{should} have for a particular set of parameters $\vectheta$ to the number we \textit{actually} used when we constructed our library. For example, if $g(M, T, A_V \mid \vectheta) = (1/2)p_{\mathrm{lib}}(M, T, A_V)$ at some particular point $(M,T,A_V)$, this means that our library has twice as many clusters in that neighbourhood as it should given the value of $\vectheta$, and thus when attempting to compute the luminosity distribution $p_{L'}(\vecL' \mid \vectheta)$ we should only count our library samples as half a cluster each, $w_j(\vectheta) = 1/2$. Note that our procedure imposes a restriction on $p_{\rm lib}(M,T, A_V)$: it must be non-zero at any point in $(M,T,A_V)$-space where $g(M, T, A_{V} \mid \vectheta)$ is non-zero for any set of parameters $\vectheta$, i.e., the support of the library must encompass the support of all candidate distributions describing the population. If this condition is not satisfied, then $w_j(\vectheta)$ diverges.

We now evaluate \autoref{eq:pl1} using \autoref{eq:obs_uncertainty} for $p(\vecL \mid \vecL',\vecsigma)$ and \autoref{eq:kde} for $p_{L'}(\vecL' \mid \vectheta)\, P_{\rm obs}(\vecL')$. This gives
\begin{eqnarray}
\lefteqn{
p_L(\vecL \mid \vectheta, \vecsigma) 
}
\nonumber \\
& \propto & 
\int \sum_{j=1}^{N_{\rm lib}} w_j(\vectheta)\,
\mathcal{N}(\vecL'\mid \vecL'_j, \vech)\, \mathcal{N}(\vecL\mid \vecL', \vecsigma) \, d\vecL'
\nonumber
\\
& \propto & 
\sum_{j=1}^{N_{\rm lib}} w_j(\vectheta)
\int \mathcal{N}(\vecL' \mid \vecL'_j,  \vech)\, \mathcal{N}(\vecL\mid \vecL', \vecsigma) \, d\vecL'
\nonumber
\\
& = & \mathcal{A}(\vectheta) \sum_{j=1}^{N_{\mathrm{lib}}} w_j(\vectheta)\, \mathcal{N}(\vecL \mid\vecL'_j, \vech'),
\label{eq:ldist}
\end{eqnarray}
where $\vech' = \sqrt{\vech^2 + \vecsigma^2}$, with the sum is computed element-wise. In the second step we use the linearity of the integration operator to exchange the sum and the integral and take the weights $w_j(\vectheta)$ out of the integral because they do not depend on $\vecL'$; in the final step we make use of the standard result for the integral of the product of normal distributions. The quantity $ \mathcal{A}(\vectheta)$ that we have added to the final line is a normalisation constant chosen to ensure that $\int p_L(\vecL \mid \vectheta, \vecsigma)\, d\vecL = 1$, and is given by
\begin{equation}
\mathcal{A}(\vectheta) = \left[\sum_{j=1}^{N_{\mathrm{lib}}} w_j(\vectheta)\right]^{-1}.
\label{eq:norm}
\end{equation}

Inserting this into \autoref{eq:likelihood1} gives the complete specification of the likelihood function,
\begin{eqnarray}
\lefteqn{p(\{\vecL_i\}, \nobs  \mid \vectheta, N_{\rm ex}, \{\vecsigma_i\}) \propto 
P_N(\nobs \mid N_{\rm ex})
}
 \qquad
\nonumber \\
& & 
\mathcal{A}(\vectheta)^{N_{\rm obs}}
\prod_{i=1}^{\nobs} 
\left[
\sum_{j=1}^{N_{\rm lib}} w_j(\vectheta)\, \mathcal{N}(\vecL_i \mid \vecL_j, \vech')
\right]
,
\label{eq:likelihood}
\end{eqnarray}
where $P_N(\nobs \mid N_{\rm ex})$ is a Poisson distribution with expectation value $N_{\rm ex}$. As noted above, since in practice we cannot constrain $N_{\rm ex}$ from observations, we can regard it as a nuisance parameter to be marginalised over. The remaining problem of determining the best-fitting parameters $\vectheta$, and exploring the shape of the posterior probability distribution in the vicinity of this maximum in order to determine uncertainties, can then be solved using any number of methods. Our implementation uses the \emcee~package \citep{foreman-mackey13a}, a Markov Chain Monte Carlo (MCMC) algorithm.

Numerical evaluation of \autoref{eq:likelihood} requires some care, because the right hand side involves a very large number of terms. A typical catalog might contain several thousand observed clusters, and the \texttt{slug} libraries we use contain $10^7$ sample clusters; thus \autoref{eq:likelihood} involves $10^{10}-10^{11}$ terms. Since any method of finding the maximum likelihood invariably involves evaluating the likelihood function hundreds of thousands of times, brute force evaluation of \autoref{eq:likelihood} is impractically slow. We can avoid this problem by noting that the normal distribution $\mathcal{N}(\vecL_i-\vecL_j \mid \vech')$ is negligibly small for most combinations of $\vecL_i$ and $\vecL_j$, because for a the great majority of clusters in the library $|(\vecL_i-\vecL_j)^2/2\vech'^2| \gg 1$. That is, only a tiny fraction of $\vecL_j$ values are near any given $\vecL_i$, and these nearby clusters completely dominate the inner sum in \autoref{eq:likelihood}. In \autoref{app:algorithm} we describe an algorithm that exploits this fact to evaluate the sum in order $\ln N_{\rm lib}$ rather than order $N_{\rm lib}$ time. Combined with openMP parallelisation over the outer product, this algorithm enables us to evaluate \autoref{eq:likelihood} for each value of $\vectheta$ and five-filter photometry in a roughly one second on a workstation, making MCMC optimisation of the fit parameters practical.

\section{Mock Catalog Tests}
\label{sec:mock}

\subsection{Generation of Mock Catalogs}

To demonstrate the capabilities of our new method, we carry out a series of tests on mock data. We generate mock star cluster data sets by running \slug~to draw a certain number of clusters from specified mass, age, and extinction distributions, and for each cluster to calculate the photometric magnitude in the \textit{Hubble Space Telescope} WFC3 filters F275W, F336W, F438W, F555W, and F814W; for shorthand below, we refer to these filters as $UV, U, B, V$, and $I$. Although our method can handle heterogenous data without difficulty, for simplicity in this demonstration of it we assume that all fields are images in these same five filters. For all the tests presented in this section, unless otherwise noted, we adopt a \citet{chabrier05a} initial mass function for the stars, we compute stellar evolution using the MIST version $1.0$ tracks for stars born rotating at 40\% of breakup \citep{dotter16a, choi16a}\footnote{The MIST models make use of the MESA stellar evolution code \citep{paxton11a, paxton13a, paxton15a}.}, and using \slug's default option (``sb99") for stellar atmospheres \citep{leitherer99a, vazquez05a}. We include extinction with an extinction law given by \slug's Milky Way extinction curve, and nebular emission using \slug's default treatment, with a ratio of nebular to stellar extinction drawn from a Gaussian distribution with a mean of 2.1 and a dispersion of 0.5, based on the empirically-determined distribution found by \citet{kreckel13a}. All tests use Solar metallicity, and all assume a constant star formation rate of $\dot{M}_* = 1$ $M_\odot$ yr$^{-1}$, with star formation at this rate having begun a time $T_{\rm sf}$ in the past, and continuing to the present. For simplicity, and because extinction is a nuisance parameter, we assume that all catalogs have an extinction distribution of the form
\begin{equation}
p({A_V}) \propto \exp\left(-\frac{A_V^2}{2\sigma_{A_V}^2}\right)
\end{equation}
with $\sigma_{A_V} = 0.5$ mag, and $A_V$ restricted to be $>0$.

\subsubsection{Mass and Age Distributions}

The different mock data sets differ only in their assumed distributions of mass and age. We consider a population of clusters born with a mass distribution
\begin{equation}
\label{eq:pM_schechter}
p(M_i) \propto M_i^{\alpha_M} \exp\left(-\frac{M_i}{M_{\rm break}}\right)
\end{equation}
above some minimum mass $M_{\rm min}$. The corresponding expectation value for the cluster mass is
\begin{equation}
\left\langle M_i \right\rangle = M_{\rm break} \frac{\Gamma\left(2+\alpha_M,M_{\rm min}/M_{\rm break}\right)}{\Gamma\left(1+\alpha_M,M_{\rm min}/M_{\rm break}\right)},
\end{equation}
where $\Gamma(a,z)$ is the incomplete $\Gamma$ function. Assuming that a fraction $f_c$ of stars are born in star clusters, the total number of clusters formed is
\begin{equation}
N_{\rm form} = f_c T_{\rm sf} \frac{\dot{M}_*}{\left\langle M_i\right\rangle}.
\end{equation}

For the age distributions, we consider two possibilities that have been advocated in the literature. Some authors \citep[e.g.,][]{fall12a, chandar15a, chandar17a} argue for mass-independent (mid) cluster disruption, whereby the probability that a given cluster survives to a particular time is independent of its mass, at least for ages below a few Gyr. In this formulation, the probability that a cluster survives to age $T$ is described by a powerlaw,
\begin{equation}
p_{\rm s,mid} =
\left\{
\begin{array}{ll}
1, & T < T_{\rm mid} \\
(T/T_{\rm mid})^{\alpha_T}, & T > T_{\rm mid}
\end{array}
\right.
\end{equation}
for $\alpha_T \leq 0$. In this case the corresponding joint distribution of cluster mass and age is
\begin{equation}
\frac{d^2 N}{dM\,dT} \propto M^{\alpha_M} \exp\left(-\frac{M}{M_{\rm break}}\right) \max\left(T, T_{\rm mid}\right)^{\alpha_T},
\label{eq:mtdist_mid}
\end{equation}
so that at ages $T > T_{\rm mid}$ we have the usual powerlaw age distribution $dN/dT \sim T^{\alpha_T}$ usually adopted in mid models. We require that this distribution only apply at ages $T > T_{\rm mid}$, and be flat at younger ages, because otherwise the distribution would diverge as $T\rightarrow 0$.\footnote{Physically the assumption that the age distribution is flat below some minimum age $T_{\rm mid}$ is expected on dynamical grounds. Even if there is a disruption mechanism that unbinds clusters on timescales below $T_{\rm mid}$, there is no way to determine from photometry that stars have become unbound until they begin to disperse, and they cannot disperse on timescales less than a cluster crossing time. Thus regardless of the nature of any physical disruption mechanism, the observed cluster age distribution must match the star formation rate distribution (i.e., must be independent of age) at times less than the typical cluster crossing time.} For this distribution,
the fraction of clusters formed that have survived to the present day, assuming $T_{\rm sf} \geq T_{\rm mid}$, is
\begin{equation}
f_{\rm s,mid} = 
\left\{
\begin{array}{ll}
\left(1 + \ln \chi \right) / \chi, & \alpha_T = -1 \\
(1/\chi^{\alpha_T}-\alpha_T/\chi)/(1-\alpha_T), & \alpha_T \neq -1
\end{array}
\right.,
\end{equation}
where $\chi = T_{\rm sf}/T_{\rm mid}$ is the number of disruption times for which star formation has been ongoing. 

The other possibility is that star clusters undergo mass-dependent disruption (mdd), as proposed for example by \citet{lamers05a} and \citet{gieles09a}. In this model clusters lose mass at a rate that varies as a powerlaw with their current mass, $dM/dT \propto M^\gmdd$ with $0 \leq \gmdd  \leq 1$, so that at age $T$ a cluster born with mass $M_i$ will have a mass
\begin{equation}
M = M_i \left[1 - \gmdd \left(\frac{M_{\rm min}}{M_i}\right)^{\gmdd} \frac{T}{\tmdd}\right]^{1/\gmdd}.
\end{equation}
If the second term in square brackets is $>1$, then the cluster is considered to have disrupted completely. Here $\tmdd$ is the timescale over which a cluster of initial mass $M_{\rm min}$ loses all its mass and disappears.\footnote{Note that the conventional choice for mdd models is to normalise to the disruption time for a cluster of initial mass $10^4$ $M_\odot$, denoted $t_4$, or an initial mass of $1$ $M_\odot$, denoted $t_0$. We have instead chosen to normalise at $M_{\rm min}$ instead, simply to avoid introducing an extra parameter. Since the disruption time is simply a powerlaw in the mass, our $\tmdd$ parameter is related to the more usual $t_4$ or $t_0$ trivially: $\tmdd = t_4 (M_{\rm min}/10^4\,M_\odot)^{\gmdd}$, and similarly for $t_0$.} In this case the distribution of present-day masses and ages is
\begin{eqnarray}
\frac{d^2 N}{dM\,dT} & \propto &  \frac{d^2 N}{dM_i\, dT} \frac{dM}{dM_i} \\
& \propto & M^{\alpha_M} \eta^{\alpha_M+1-\gmdd}
\exp\left(-\eta \frac{M}{M_{\rm break}}\right)
\label{eq:mtdist_mdd}
\end{eqnarray}
where
\begin{equation}
\eta(M,T) \equiv  \left[1+\gmdd\left(\frac{M_{\rm min}}{M}\right)^\gmdd \frac{T}{\tmdd}\right]^{1/\gmdd}
\end{equation}
is the ratio of the initial mass to the present mass for a cluster of present-day mass $M$ and age $T$. Since a cluster born of age $T$ must have been formed with a mass larger than $M_{\rm s,min} = M_{\rm min} (\gmdd T/\tmdd)^{1/\gmdd}$ to have survived, the fraction of clusters of age $T$ that have survived mass-dependent disruption is
\begin{eqnarray}
f_{\rm s,mdd}(T) & = & \frac{\int_{\max(M_{\rm s,min}(T),M_{\rm min})}^\infty M^{\alpha_M} e^{-\frac{M}{M_{\rm break}}}\, dM}{\int_{M_{\rm min}}^\infty M^{\alpha_M} e^{-\frac{M}{M_{\rm break}}}\, dM} 
\\
& = &
\frac{\Gamma\left(1+\alpha_M, \frac{\max(M_{\rm s,min}(T),M_{\rm min})}{M_{\rm break}}\right)}{\Gamma\left(1+\alpha_M, \frac{M_{\rm min}}{M_{\rm break}}\right)}.
\end{eqnarray}
Averaged over all ages, the total fraction of surviving clusters is
\begin{equation}
f_{\rm s,mdd} = \frac{1}{T_{\rm sf}} \int_0^{T_{\rm sf}} f_{\rm s,mdd}(T)\, dT.
\label{eq:fsmdd}
\end{equation}
The integral cannot be evaluated in closed form, but is trivial to evaluate numerically for any specified set of parameters.

\subsubsection{Mock Catalogs}

\begin{table*}
\centerline{
\begin{tabular}{llcccccc}
\hline
\multicolumn{2}{c}{Parameter} & \texttt{Powerlaw} & \texttt{Truncated} & \texttt{MDD} & \texttt{DoubleErr} & \texttt{LibMismatch} & \texttt{CompMismatch}  \\
\hline
$N_{\rm obs}$ & & 5629 & 5167 & 5479 & 5255 & 5088 & 5093 \\[2ex]
\multirow{3}{*}{mid / mdd} 
& True & mid & mid & mdd & mid & mid & mid \\
& $w(\mbox{mid})$ & 1.0 & 1.0 & $<10^{-10}$ & 1.0 & 1.0 & 1.0 \\
& $w(\mbox{mdd})$ & $<10^{-10}$ & $<10^{-10}$ & 1.0 &  $<10^{-10}$ & $<10^{-10}$ & $<10^{-10}$ \\ [2ex]
\multirow{2}{*}{$\alpha_M$} 
& True & $-2$ & $-2$ & $-2$ & $-2$ & $-2$ & $-2$ \\
& Fit & $-2.00_{-0.021}^{+0.018}$ & $-2.01_{-0.033}^{+0.028}$ & $-2.05_{-0.039}^{+0.019}$ & $-2.00_{-0.021}^{+0.019}$ & $-1.91_{-0.033}^{+0.030}$ & $-1.97_{-0.023}^{+0.020}$ \\ [2ex]
\multirow{2}{*}{$\log (M_{\rm break}/M_\odot)$}
& True & $6.5$ & $5.0$ & $5.0$ & $5.0$ & 5.0 & 5.0 \\
& Fit & $6.36_{-0.39}^{+0.51}$ & $4.90_{-0.12}^{+0.17}$ & $5.20_{-0.09}^{+0.16}$ & $4.90_{-0.08}^{+0.10}$ & $4.76_{-0.08}^{+0.10}$ & $4.96_{-0.11}^{+0.11}$ \\ [2ex]
\multirow{2}{*}{$\alpha_T$}
& True & $-1$ & $-1$ & -- & $-1$ & $-1$ & $-1$ \\
& Fit & $-1.02_{-0.022}^{+0.021}$ & $-1.01_{-0.056}^{+0.071}$ & -- & $-0.97_{-0.038}^{+0.040}$ & $-0.95_{-0.035}^{+0.040}$ & $-1.04_{-0.040}^{+0.042}$ \\ [2ex]
\multirow{2}{*}{$\log(T_{\rm mid}/\mathrm{yr})$}
& True & 6.5 & 8.0 & -- & 8.0 & 8.0 & 8.0 \\
& Fit & $6.49_{-0.031}^{+0.029}$ & $8.00_{-0.050}^{+0.047}$ & -- & $7.96_{-0.030}^{+0.031}$ & $7.90_{-0.041}^{+0.042}$ & $7.97_{-0.032}^{+0.035}$  \\ [2ex]
\multirow{2}{*}{$\gmdd$}
& True & -- & -- & $0.65$ & -- & -- & -- \\
& Fit & -- & -- & $0.61_{-0.036}^{+0.030}$ & -- & -- & -- \\ [2ex]
\multirow{2}{*}{$\log(\tmdd/\mathrm{yr})$}
& True & -- & -- & 6.98 & -- & -- & -- \\
& Fit & -- & -- & $7.05_{-0.049}^{+0.134}$ & -- & -- & -- \\ [2ex]
\hline
\end{tabular}
}
\caption{
\label{tab:catalogs}
Parameters of mock catalogs, and results of fits to these parameters using \slug. The first row lists the number of observed clusters in each mock catalog. The second row specifies whether the catalog was generated using mass-independent or mass-dependent disruption (mid or mdd), and the values $w(\mathrm{mid})$ and $w(\mathrm{mdd})$ that we report are the Akaike weights of the mid and mdd models as determined from our MCMC fits; see main text for details. For all other parameters, we give the true value used in generating the catalog, and we list fit values in the form $\left(q_{50}\right)^{+(q_{84}-q_{50})}_{-(q_{50}-q_{16})}$ where $q_{N}$ is the $N$th percentile estimate for $q$. Thus the value reported is the 50th percentile, and the $+$ and $-$ error range indicates the range from the 16th to the 84th percentile. While we calculate fit parameters for both mid and mdd models for each mock catalog, in the table above we report the fits only for whichever of the two models has the higher Akaike weight.
}
\end{table*}

Our joint mass-age distribution is fully characterised by a choice to use mass-dependent or mass-independent disruption, and by six parameters:  $M_{\rm min}$, $T_{\rm sf}$, $\alpha_M$, $M_{\rm break}$, and either $\alpha_T$ and $T_{\rm mid}$ (for mid models) or $\gmdd$ and $\tmdd$ (for mdd models). The parameters $M_{\rm min}$ and $T_{\rm sf}$ cannot generally be determined from observations of the type we are considering because clusters near the minimum mass or maximum age are invariably too dim to observe; they enter the problem only by changing the total number of clusters in the catalog. Since this effect is degenerate with changes to the value of $f_c$ (the fraction of stars formed in clusters) or $\dot{M}_*$ (the total star formation rate), we simply set $M_{\rm min} = 100$ $M_\odot$ and $T_{\rm sf} = 10$ Gyr for all catalogs, and do not explore variations in these parameters further.  For the remaining parameters, we consider three mock catalogs, whose parameters are summarised in \autoref{tab:catalogs}, that illustrate different possible combinations of them. We tune our parameters so that each catalog produces a comparably-sized sample of observable clusters, with the size chosen to be about the size of the catalog for NGC 628 presented by \citet{adamo17a}, which contains approximately 3,700 clusters (though not all are visually confirmed). The cases are:

\begin{itemize}
\item \texttt{Powerlaw}: this case uses a mock catalog with clusters drawn from a distribution similar to that proposed by, e.g., \citet{fall12a}, \citet{chandar15a}, and \citet{chandar17a}, whereby the mass function is a pure powerlaw whose upper limit set set only by size of sample effects, and disruption is mass independent. Specifically, we adopt for this catalog $\alpha_M = -2$, $\alpha_T = -1$, $M_{\rm break} = 10^{6.5}$ $M_\odot$,  and $T_{\rm mid} = 10^{6.5}$ yr. Note that for this choice of $M_{\rm break}$ the expected number of clusters that form with $M > M_{\rm break}$ is $\ll 1$, so the mass distribution is effectively a pure powerlaw truncated only by finite sample size. Also adopting $f_c = 1$ for this case, the expected number of clusters surviving to the present day is 
\begin{equation}
N = N_{\rm form} f_{\rm s,mid} = 2.93\times 10^4.
\end{equation}
We therefore draw this number clusters for the mock catalog. 

\item \texttt{Truncated}: this case is similar to the results obtained by \citet{fouesneau14a} and \citet{johnson17b} for Andromeda. Disruption is mass-independent, but no disruption occurs until ages above 100 Myr, i.e.,  $T_{\rm mid} = 10^8$ yr. The mass function is truncated at a lower mass, $M_{\rm break} = 10^5$ $M_\odot$. Finally, only 10\% of stars form in clusters, so $f_c = 0.1$. All other parameters are the same as for \texttt{Powerlaw}. For this distribution the mean cluster mass is $\langle M \rangle = 638$ $M_\odot$ and the number of surviving clusters is $N = 8.78\times 10^4$.

\item \texttt{Mass-dependent disruption (MDD)}: this catalog uses parameters chosen to be similar to those obtained by \citet{adamo17a}. For this case we use mass-dependent disruption, with $\gmdd = 0.65$ and $\tmdd = 9.5$ Myr\footnote{Note that $\tmdd=9.5$ Myr is equivalent to \citeauthor{adamo17a}'s parameter $t_4 = 190$ Myr.}. The mass distribution at birth is the same as for \texttt{Truncated}, i.e.,  $\alpha_M = -2$ and $M_{\rm break} = 10^5$ $M_\odot$. We take $f_c = 0.3$ in this case, so the total number of clustered formed is $N_{\rm form} = 4.71\times 10^{6}$. Using \autoref{eq:fsmdd}, the fraction of clusters that survive mass-dependent disruption is $f_{\rm s,mdd} = 0.00370$, so the expected number of clusters at the present day is  $N = 1.74\times 10^4$. 

\item \texttt{DoubleErr}: this catalog is identical to \texttt{Truncated}, except that the assumed photometric errors added to the true cluster luminosities are twice as large -- see below. Its purpose is to test how our results depend on the size of the photometric error.

\item \texttt{LibMismatch}: this catalog is identical to \texttt{Truncated} in its parameters, but instead of generating the mock catalog using MIST models for stellar evolution and a Milky Way extinction curve, we generate them using \slug's Padova tracks \citep{girardi00a} and starburst attenuation curve. The goal of this catalog is to test the robustness of our method in a case where the models for stellar evolution and dust are not a perfect match to the underlying data.

\item \texttt{CompMismatch}: this catalog is identical to \texttt{Truncated} in its parameters, but uses a different completeness function (see below). The goal of this catalog is to test how our method behaves when our estimated completeness function is not exactly correct.

\end{itemize}

\subsubsection{Completeness and Photometric Error}
\label{sssec:completeness}

\begin{figure}
\includegraphics[width=\columnwidth]{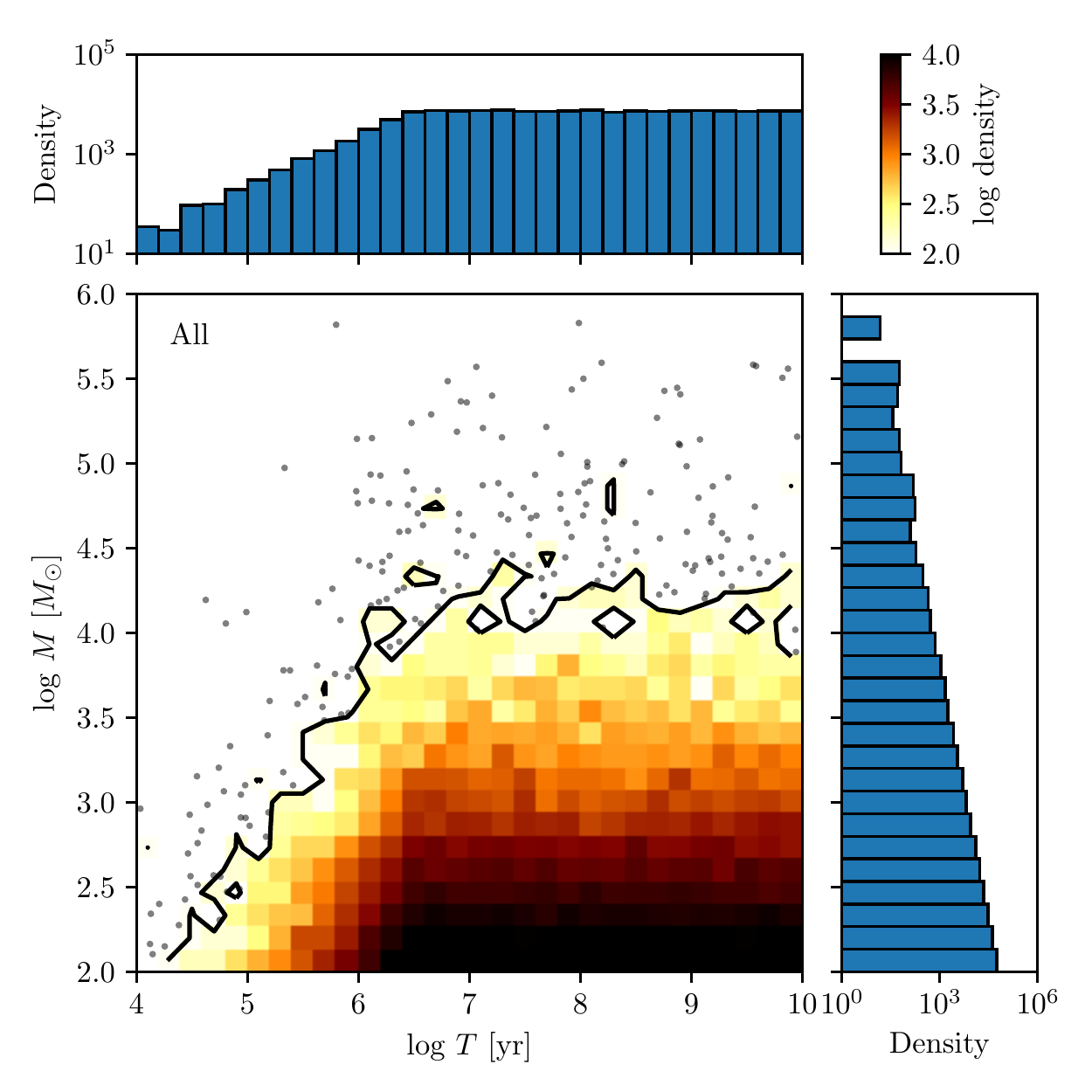}
\includegraphics[width=\columnwidth]{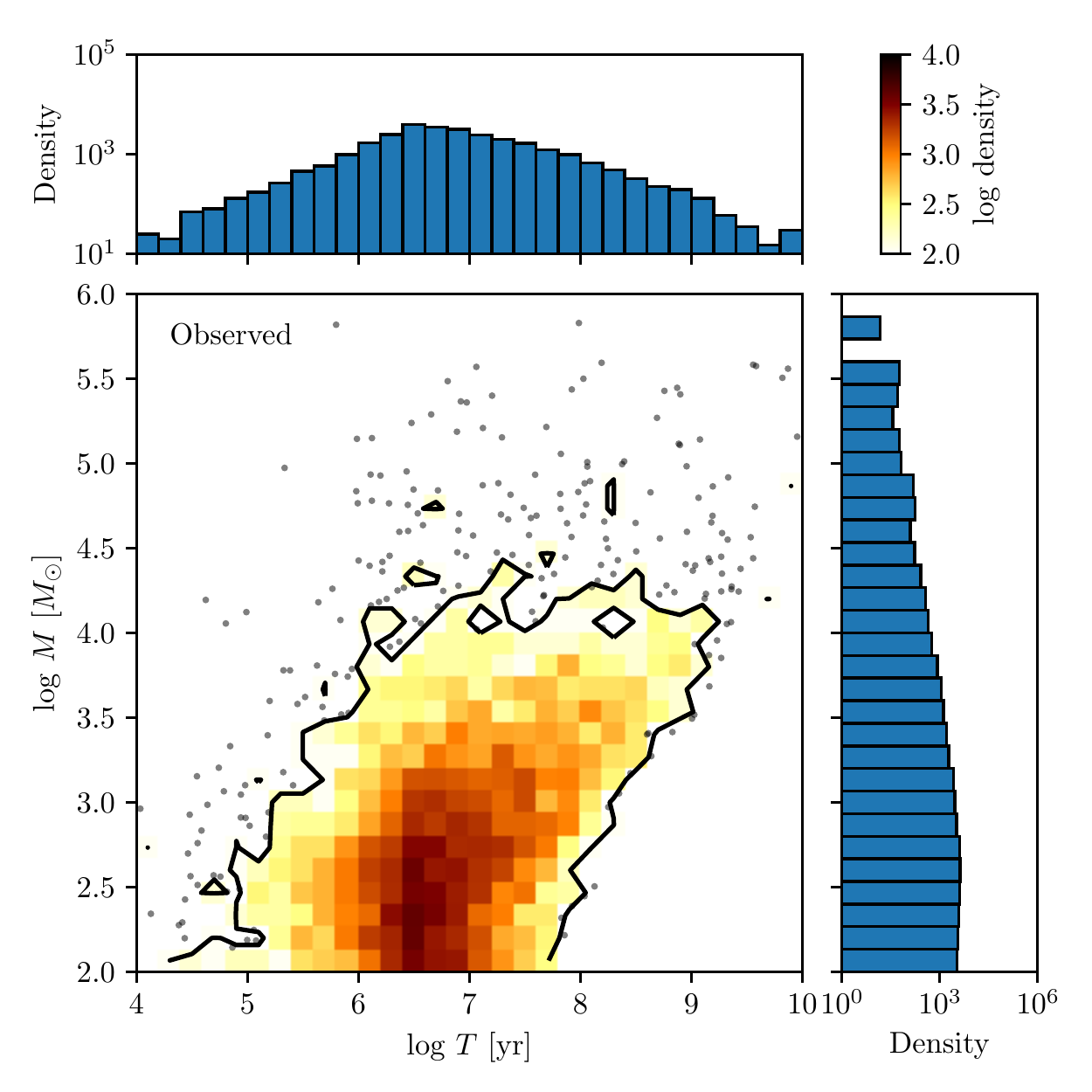}
\caption{
\label{fig:mock_phys}
Distribution of clusters in the \texttt{Powerlaw} mock catalog. The top set of panels shows the distribution of all clusters, while the bottom set shows the distribution of those clusters that are observed. In each set of panels, the central one shows the density of clusters (in clusters per $\mbox{dex}^2$), as indicated by the colour bar; points mark individual clusters in sparsely-populated regions. The enclosing contour corresponds to a number of clusters per $\mbox{dex}^2$ equal to the lowest value in the colour bar. Above and to the right of the central panel we show one-dimensional histograms of the mass and age distributions, in units of clusters per dex (i.e., the quantities plotted are $dN/d\log M$ and $dN/d\log T$).
}
\end{figure}

\begin{figure}
\includegraphics[width=\columnwidth]{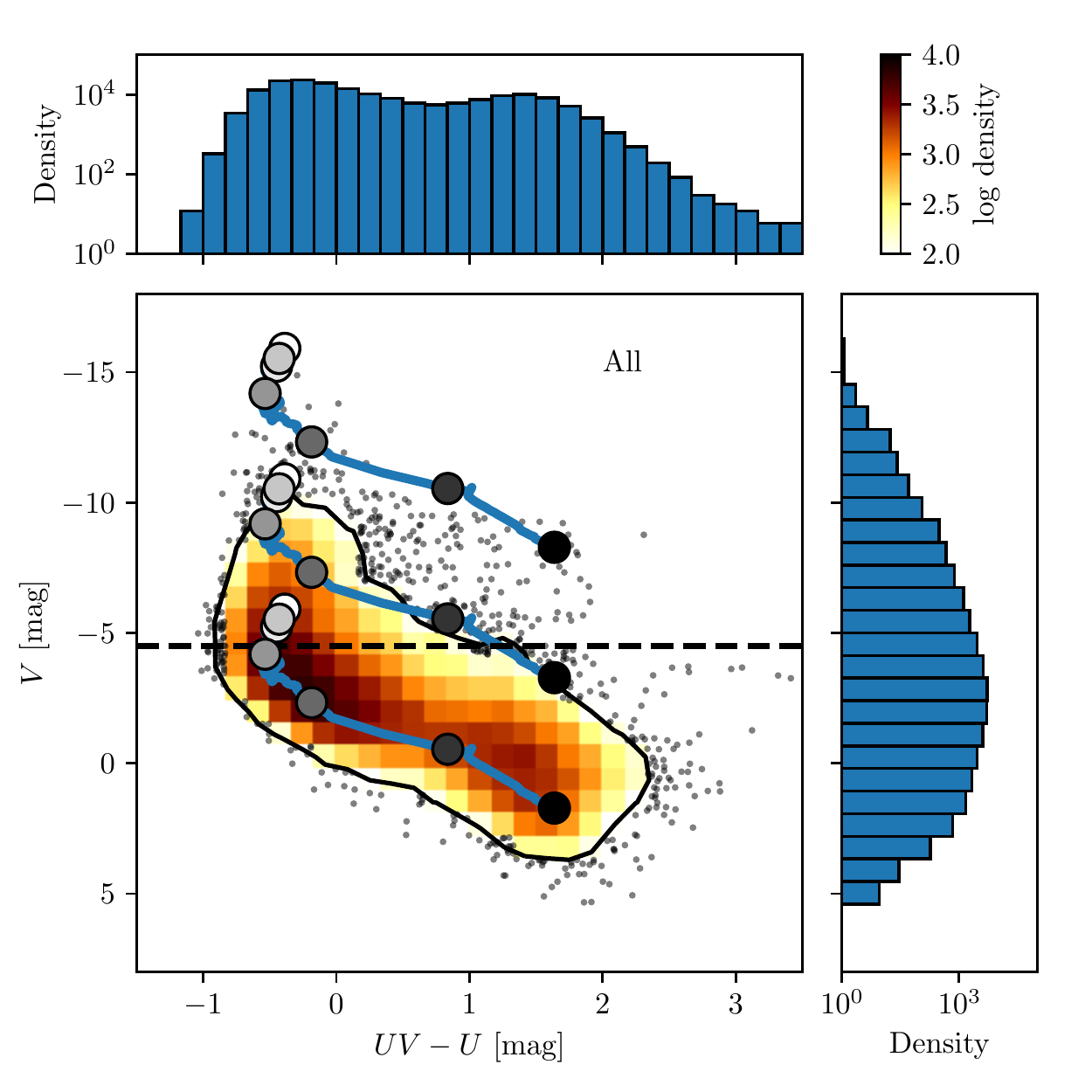}
\includegraphics[width=\columnwidth]{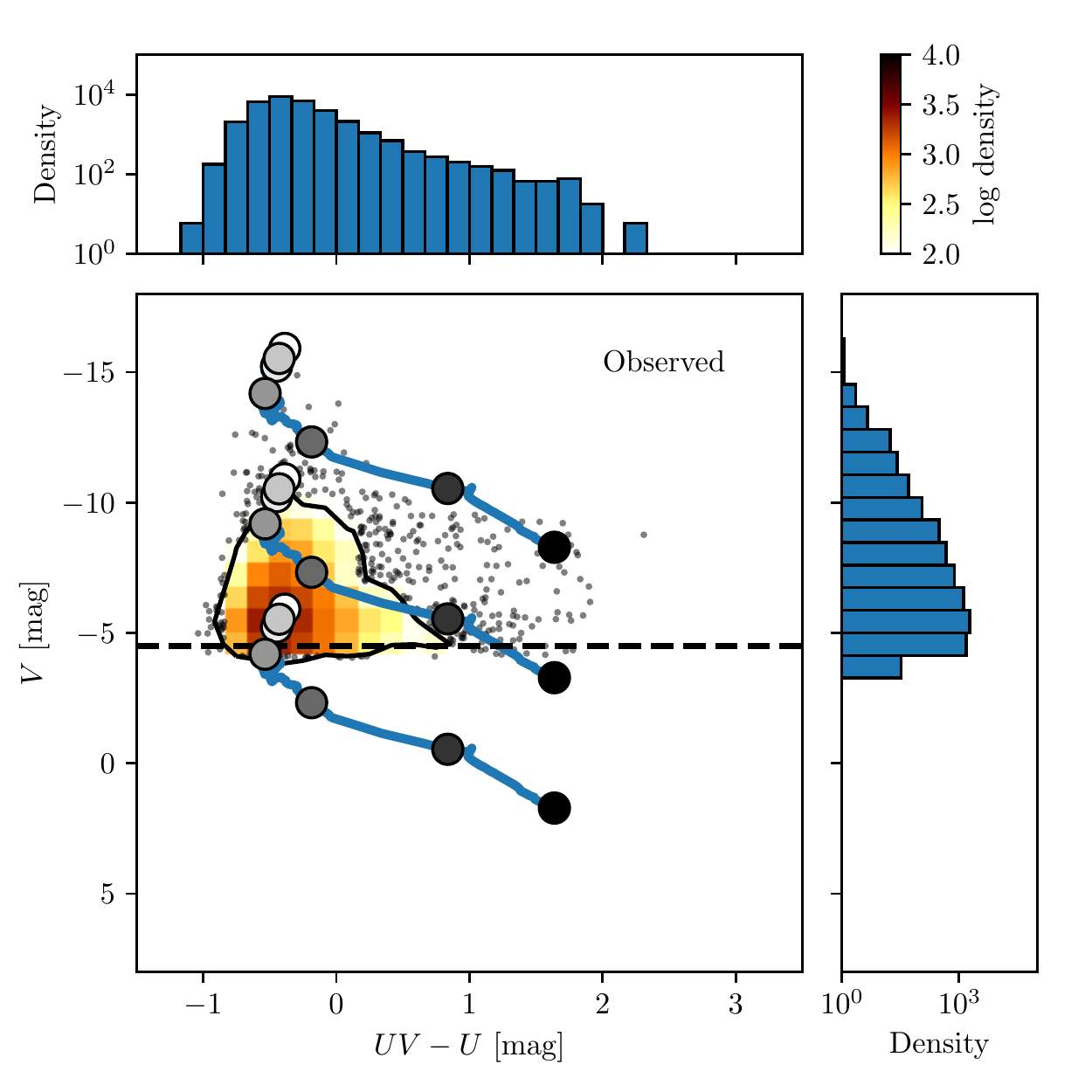}
\caption{
\label{fig:mock_colourmag}
Same as \autoref{fig:mock_phys}, except that now we show the distribution of clusters in colour and magnitude rather than in their physical properties. The blue lines with large gray points show evolutionary tracks for unexctincted clusters with fully sampled (i.e., non-stochastic) stellar populations over the age range from $10^5 - 10^{10}$ yr. From top to bottom, the lines correspond to cluster masses of $10^6$, $10^4$, and $10^2$ $M_\odot$. The points are logarithmically-spaced in age from $10^5$ yr (lightest) to $10^{10}$ yr (darkest) at intervals of 1 dex. The dashed black line indicates the 50\% completeness limit.
}
\end{figure}

\begin{figure}
\includegraphics[width=\columnwidth]{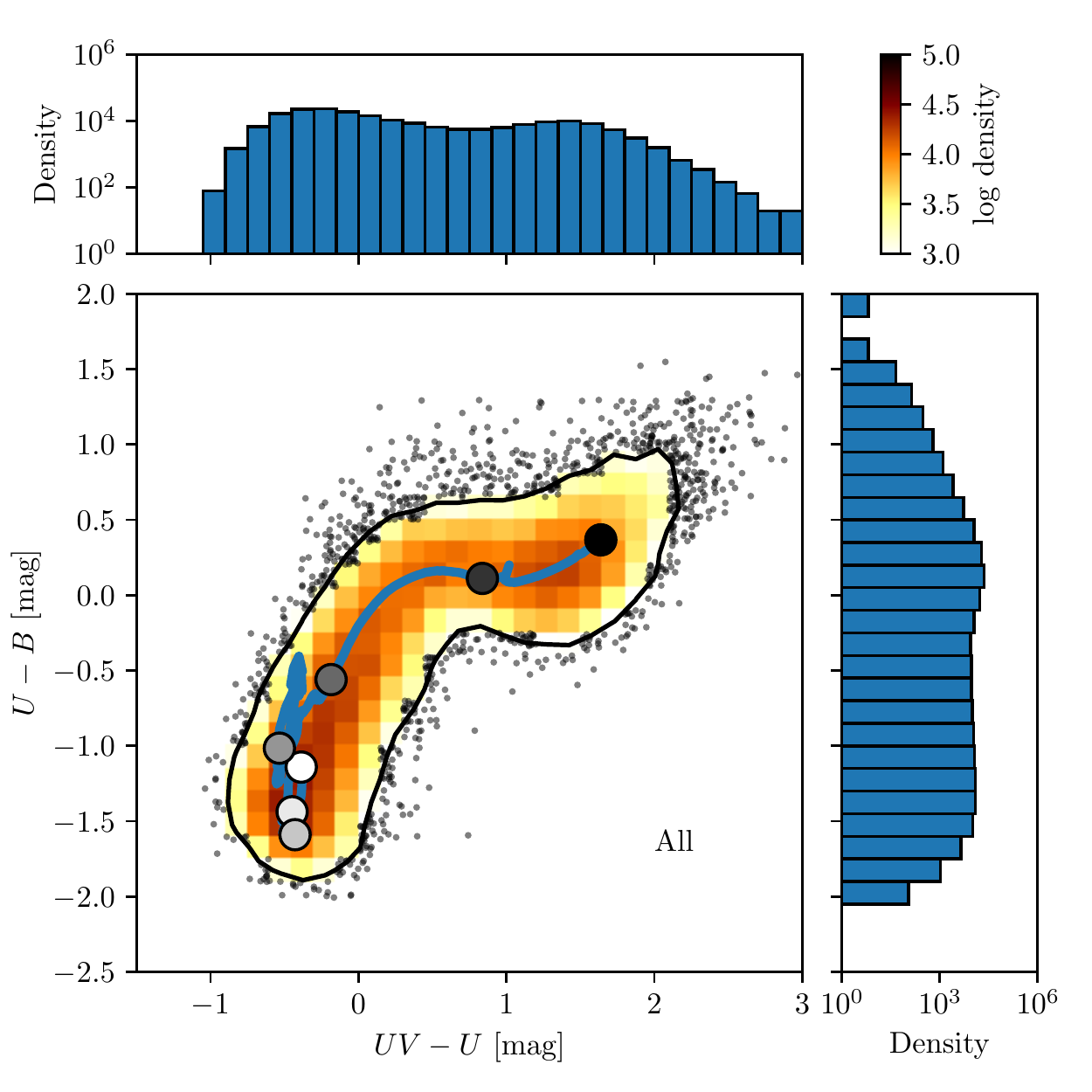}
\includegraphics[width=\columnwidth]{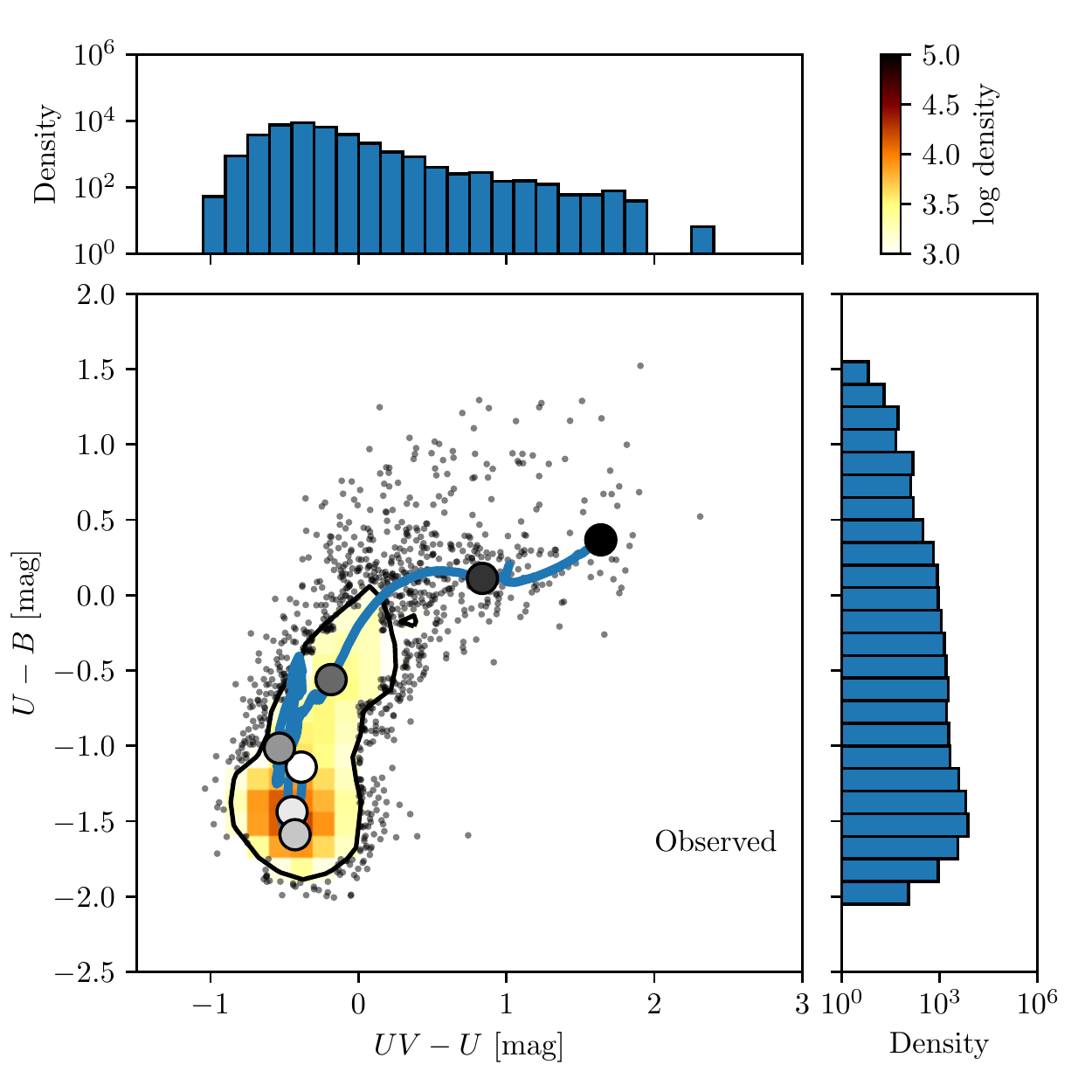}
\caption{
\label{fig:mock_colourcolour}
Same as \autoref{fig:mock_colourmag}, except that now we show the distribution in colour-colour space rather than colour-magnitude space.
}
\end{figure}

To test the effects of observational completeness and photometric error, and show how our method copes with them, we next add noise to our mock catalogs, and apply completeness cuts to them. We set the photometric noise level for all catalogs except \texttt{DoubleErr} to $0.1$ mag in all bands\footnote{Here and throughout we use Vega magnitudes.}, based on typical levels of photometric accuracy in recent large surveys such as LEGUS \citep{adamo17a}. To test the sensitivity of our results to the noise level, for \texttt{DoubleErr} we set the noise level to $0.2$ mag instead, comparable to the poorest levels of accuracy in LEGUS. For either noise level, we generate the observed magnitudes of all clusters by taking the true magnitudes in each band calculated by \texttt{slug} and adding a random offset drawn from a Gaussian distribution with a dispersion 0.1 or 0.2 mag, as appropriate for that catalog.

We next apply a completeness cut, using a completeness function comparable to that obtained by \citet{adamo17a} for the galaxy NGC 628 based on their mock cluster tests. Specifically, for all the catalogs except \texttt{CompMistmatch}, we take the probability that a given cluster makes it into the catalog to be 100\% for $V \leq -5$ mag, 0 for $V \geq -4$ mag, and linearly varying between these two limits for $-5\mbox{ mag} < V < -4\mbox{ mag}$, i.e., $P_{\rm obs} = -V - 4$ mag. For \texttt{CompMismatch}, we instead use a completeness function that is 100\% for $V \leq -4.75$ mag, 0 for $V \geq -4$ mag, and varies in between these two limits as $P_{\rm obs} = [(-V - 4)/0.75]^2$. For each cluster we assign a flag of ``observed" or ``not observed" based on the $V$ magnitude; for clusters in the partially complete range, we randomly assign them one flag or the other with probability $P_{\rm obs}(V)$. This process yields a list of $\approx 5000$ observed clusters for of our mock catalogs; the exact number in each case is given in \autoref{tab:catalogs}.

As an example of the effects of noise and the completeness cut, \autoref{fig:mock_phys} shows the true distribution of cluster physical properties in the \texttt{Powerlaw} catalog, and the corresponding distribution for those clusters flagged as observed. \autoref{fig:mock_colourmag} and \autoref{fig:mock_colourcolour} show the corresponding observed colour-magnitude and colour-colour diagrams. As the plots shows, observational completeness truncates both the mass and age distributions, and does so in a way that is correlated -- clusters are more likely to remain in the catalog if they are either young or massive, and are mostly removed if they are old and low mass. However, the cutoff imposed by observational limits is not sharp in either mass or age due to the effects of stochastic sampling, varying extinction, and partial completeness. For example, for a fully-sampled (i.e., non-stochastic), unextincted stellar population with a mass of $300$ $M_\odot$, the ages corresponding to 100\%, 50\%, and 0\% completeness ($V = -5, -4.5$, and $-4$ mag, respectively) are 12.0, 19.1, and 53.7 Myr, respectively. However, in our \texttt{Powerlaw} mock catalog, we find that there are 13 observed clusters with mass $<300$ $M_\odot$ at ages above 53.7 Myr, and 388 non-observed clusters larger than $300$ $M_\odot$ with ages below 12.0 Myr.

The conventional means of avoiding this complication in analysing cluster populations is to impose fairly severe cuts on the data so as to ensure that the sample that is retained is well within the zone of completeness, and massive enough to be relatively unaffected by stochasticity. However, this approach is undesirable because it both discards much of the available information and restricts the range of applicability of the resulting fits. For example, our \texttt{Powerlaw} mock catalog contains 5,629 observed clusters. Discarding all clusters with $V > -6$ mag, as is done for example in \citet{adamo17a}, leaves only 2,445, and thus amounts to throwing out more than half the data. Any fits to the remaining data could be used to constrain the distributions of mass and age only for masses above $10^4$ $M_\odot$ and ages below 1 Gyr (or masses above $10^{3.7}$ $M_\odot$ and ages below $200$ Myr, the cuts used in \citealt{adamo17a}), since only in this mass-age range are the data reasonably complete.

\subsection{Libraries}

To analyse the mock catalogs we require libraries of model clusters. To produce these we use \texttt{slug} to simulate $10^7$ clusters using the same combination of tracks, atmospheres, and metallicity as for all the mock catalogs except \texttt{LibMismatch}. Cluster masses, ages, and extinctions in the library are drawn randomly from the distribution
\begin{equation}
p_{\rm lib}(M, T, A_V) \propto p_M(M) p_T(T) p_{A_V}(A_V)
\end{equation}
with
\begin{eqnarray}
p_M(M)& \propto &
\left\{
\begin{array}{ll}
\left(\frac{M}{10^5\,M_\odot}\right)^{-1}, \quad & 10^2 < \frac{M}{M_\odot} \leq 10^5 \\
\left(\frac{M}{10^5\,M_\odot}\right)^{-2}, & 10^5 < \frac{M}{M_\odot} \leq 10^7 \\
\end{array}
\right.
\\
p_T(T) & \propto & \frac{1}{T}, \quad 10^5\,{\rm yr} < T < 1.5\times 10^{10}\,{\rm yr}
\\
p_{A_V}(A_V) & \propto & \mbox{const}, \quad 0 < A_V < 3\,{\rm mag}.
\end{eqnarray}
This sampling is chosen to maximise the density of samples at younger masses and ages where there is a larger amount of stochastic variation in cluster colour and luminosity; we remind the reader that the sampling density is explicitly accounted for in \autoref{eq:wgts}, and thus this choice affects the result only insofar as it provides a better or worse sampling of the underlying distribution. All other details of the sampling procedure are identical to that used in \citet{krumholz15c}, and we refer readers to that paper for full details.

We adopt a bandwidth of $\vech = 0.05$ dex in the physical dimensions and $\vech = 0.05$ mag in the photometric directions. Note that the value of $\vech$ enters the calculation only through evaluation of the likelihood function, \autoref{eq:likelihood}. This means that the value of $\vech$ in the physical dimensions has no effect on the results, because the likelihood function only makes use of the photometric dimensions. In the photometric dimensions, $\vech$ enters only in quadrature sum with the uncertainties $\vecsigma$, and thus the value of $\vech$ does not affect the results as long as $h$ is significantly smaller than $\sigma$ in all dimensions. Our choices satisfy this condition, and in general the condition can always be satisfied as long as the library is large enough to allow a choice of $\vech$ satisfying this condition. For a more detailed exploration of values of $\vech$ and the density of sampling points in the library, we refer readers to \citet{krumholz15c}.

\subsection{Analysis of the Mock Catalogs}
\label{ssec:mock_analysis}

\begin{figure}
\centerline{
\includegraphics[width=\columnwidth]{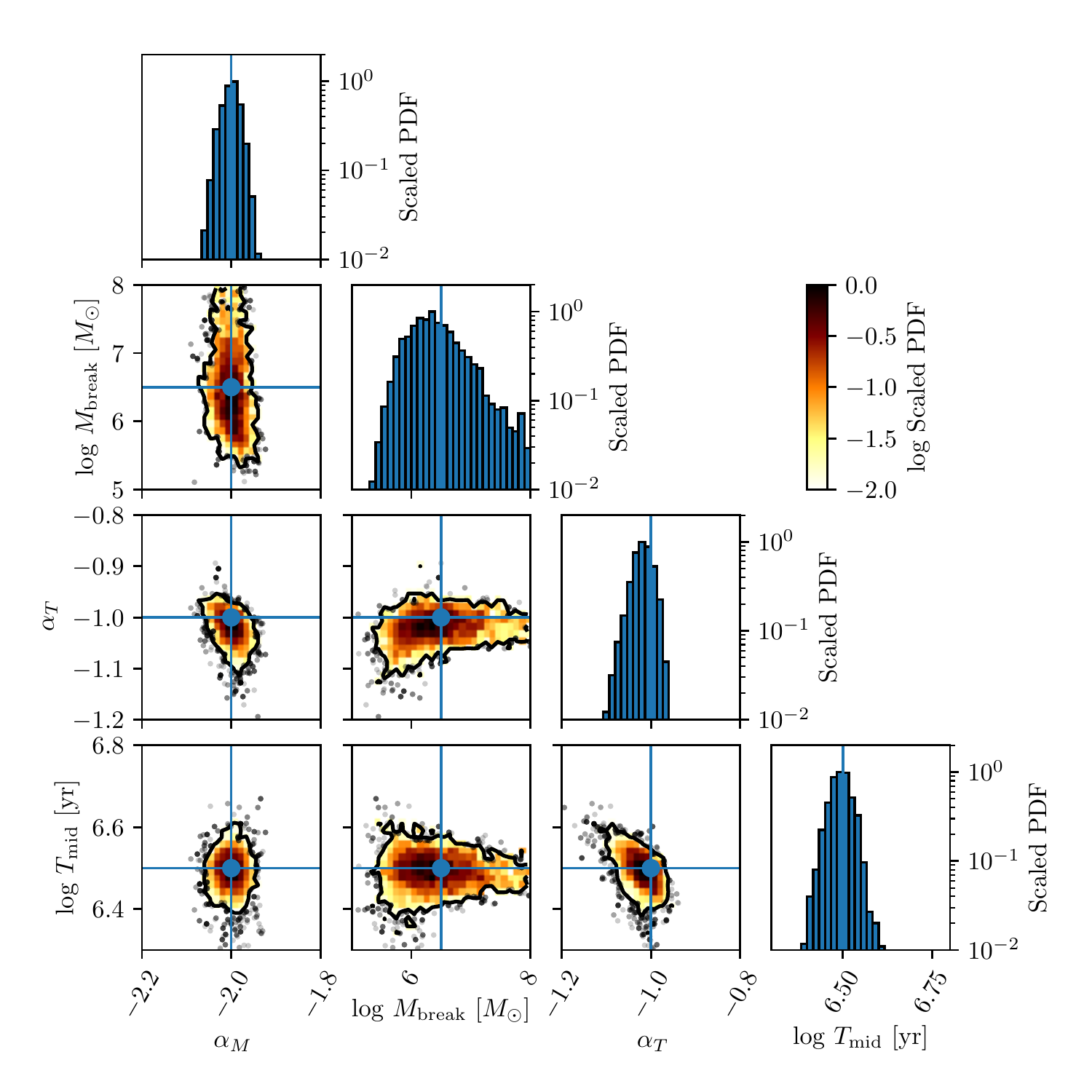}
}
\caption{
\label{fig:posteriors_powerlaw}
Corner plot showing the 1D and 2D histograms of the posterior PDFs of the parameters $\alpha_M$, $\log M_{\rm break}$, $\alpha_T$, and $\log\, T_{\rm mid}$ for the \texttt{Powerlaw} mock catalog, as determined by MCMC optimisation. Nuisance parameters describing the dust extinction distribution have been omitted. Blue histograms show 1D marginal PDFs for each parameter; red-coloured heat maps show 2D probability densities on a logarithmic scale, with all panels normalised to have a maximum of unity. The contour corresponds to a probability density of $10^{-2}$ on this scale, and scattered points show individual MCMC samples outside this contour. Blue lines and points indicate the true values for the input \texttt{Powerlaw} catalog.
}
\end{figure}

\begin{figure}
\centerline{
\includegraphics[width=\columnwidth]{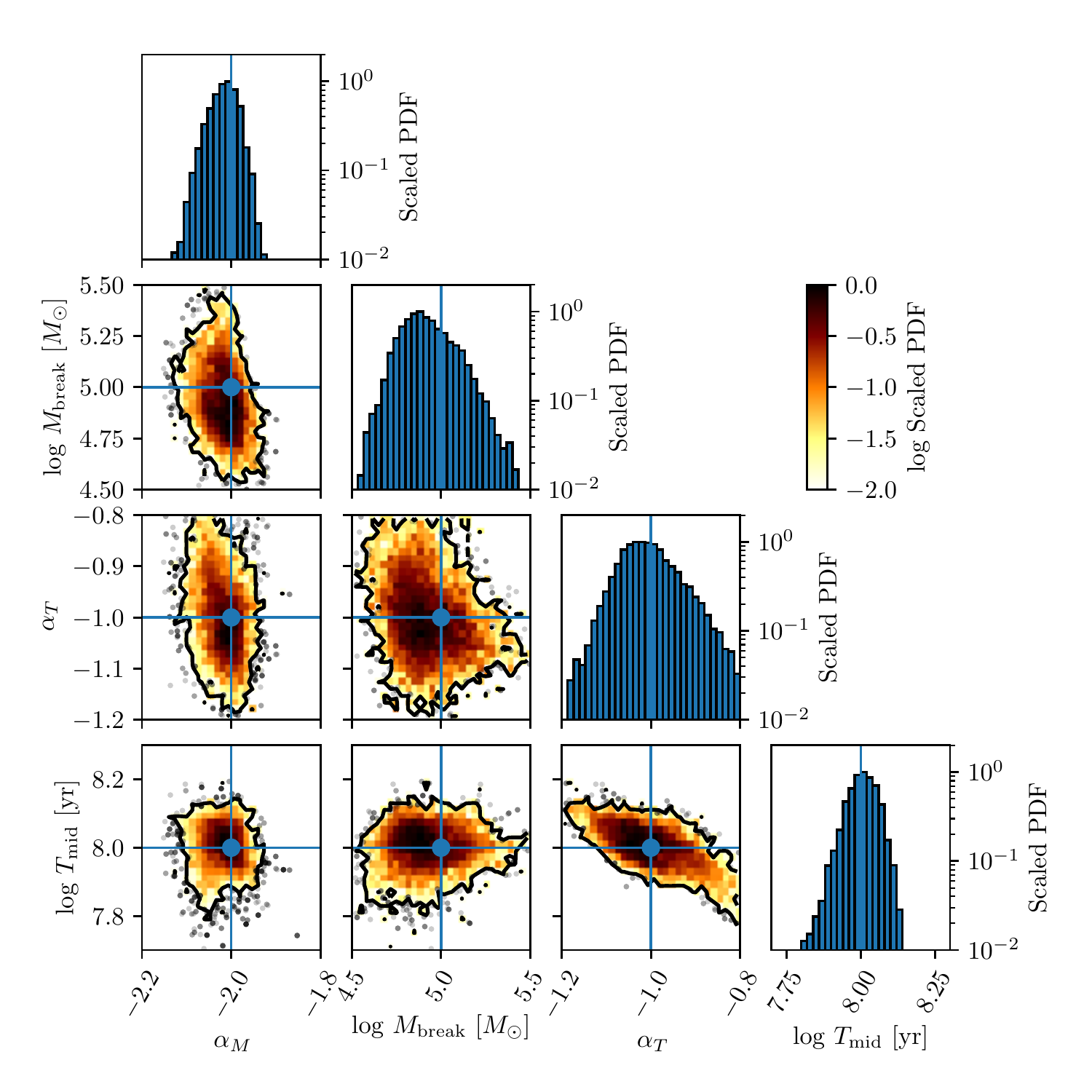}
}
\caption{
\label{fig:posteriors_truncated}
Same as \autoref{fig:posteriors_powerlaw} for the \texttt{Truncated} catalog.
}
\end{figure}

\begin{figure}
\centerline{
\includegraphics[width=\columnwidth]{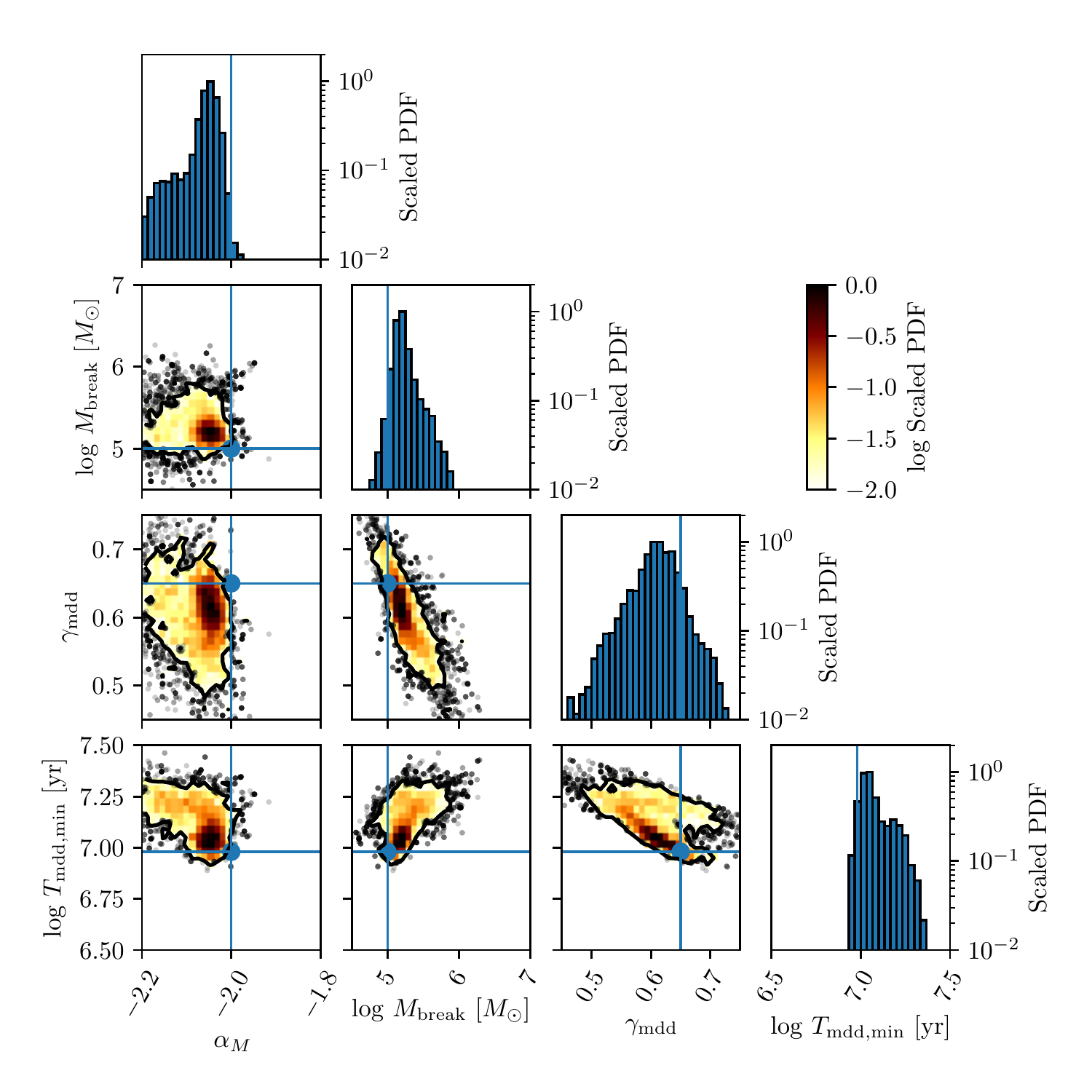}
}
\caption{
\label{fig:posteriors_mdd}
Same as \autoref{fig:posteriors_powerlaw} for the \texttt{MDD} catalog. However, note that the last two columns plotted (for the quantities $\gamma_{\rm mdd}$ and $T_{\mathrm{mdd,min}}$) are different than in \autoref{fig:posteriors_powerlaw}, since the AIC indicates that the mdd model is a better match to this catalog than the mid model.
}
\end{figure}

We first focus on the \texttt{Powerlaw}, \texttt{Truncated}, and \texttt{Mdd} catalogs, which have uniform tracks, completeness, and errors, and where the only differences are between the parameters describing the cluster distribution; we discuss the remaining cases in the next section.
For each mock catalog we consider proposed distributions of cluster mass and age following the functional forms given by \autoref{eq:mtdist_mid} and \autoref{eq:mtdist_mdd}; the free parameters are $\alpha_M$, $M_{\rm break}$, and either $\alpha_T$ and $T_{\rm mid}$ (for \autoref{eq:mtdist_mid}) or $\gmdd$, and $\tmdd$ (for \autoref{eq:mtdist_mdd}). We leave $M_{\rm min}$ and $T_{\rm sf}$ fixed as above. Our priors on $\alpha_M$, $\alpha_T$, and $\gmdd$ are flat. For $\log M_{\rm break}$, our priors are restricted by the range of masses sampled in our library, and thus we adopt a prior that is flat in $\log (M_{\rm break}/M_\odot)$ from $2-7$. For the same reason we adopt priors on $\log (T_{\rm mid}/\mbox{yr})$, and $\log(\tmdd/\mbox{yr})$ that are flat from $5 - 10.17$.

Since we cannot assume that we know the functional form of the dust extinction, we choose to parameterise it with a simple piecewise-linear form over the range $0-3$ mag in our library. Specifically, we define $A_{V,i} = i \Delta A_V$ mag for $i = 0 \ldots N$, where $\Delta A_V = 3/N$ mag, and our linear fit breaks the range from 0 to 3 mag into $N$ intervals. For an extinction $A_V$ in the range $[A_{V,i},A_{V,i+1})$ we set $p(A_V) = p_{A_V,i} + (p_{A_V,i+1} - p_{A_V,i}) (A_V - A_{V,i})/\Delta A_V$. The values of $p_{A_V,i}$ for $i=0\ldots N-1$, representing the values of the extinction PDF at points $A_{V,i}$, are free parameters to be fit, while the value of $p_{A_V,N}$ is fixed by the requirement that $\int p(A_V)\,dA_V = 1$. In the experiments we present here we adopt $N=6$, corresponding to breaking the extinction PDF into bins 0.5 mag wide, but our code leaves this as a free parameter to be set at run time. We adopt priors that are flat in $\log p_{A_V,i}$ for $i=0 \ldots N-1$, subject to the constraint that $p_{A_V,N} > 0$.

We carry out the optimisation of the parameters for both the mid and mdd cases using 100 walkers and 500 iterations; visual inspection shows that the distribution of walkers stabilises after $\sim 150$ iterations, so we discard the first 200 iterations and derive the posterior PDFs from the remainder. For our sample catalogs, each MCMC calculation requires $\approx 12$ hours on a workstation. To decide whether the mid or mdd model provides a better fit to each data set, we compute the Akaike information criterion (AIC) for each model \citep[e.g.,][]{sharma17a}. Specifically, for the mid and mdd cases we find the largest value of the likelihood function $p\left(\{\vecL_i\}, N_{\rm obs} \mid \vectheta, N_{\rm ex}, \{\vecsigma_i\}\right)$ returned by any of the MCMC sample points, which we denote $\hat{\mathcal{L}}$, and compute
\begin{equation}
\mbox{AIC}_{\rm (mid,mdd)} = 2 k - 2 \ln \hat{\mathcal{L}}_{\rm (mid,mdd)},
\end{equation}
where $k=11$ is the number of parameters for both the mid and mdd models: 6 parameters to describe the dust extinction distribution, 4 to described the joint mass-time distribution, and 1 to describe the number of clusters $N_{\rm ex}$. The corresponding Akaike weight for the mid model,
\begin{eqnarray}
w(\mathrm{mid}) & = & \frac{e^{-\Delta_{\rm mid}/2}}{e^{-\Delta_{\rm mid}/2} + e^{-\Delta_{\rm mdd}/2}} \\
\Delta_{\rm (mid,mdd)} & = & \mbox{AIC}_{\rm (mid,mdd)} - \min(\mbox{AIC}_{\rm mid}, \mbox{AIC}_{\rm mdd}),
\end{eqnarray}
gives the probability that the mid model is the better fit to the data. Note that this method automatically marginalises over the unknown dust distribution.

We report values of $w({\rm mid})$ (and the analogously-determined $w({\rm mdd}) = 1 - w({\rm mid})$), and posteriors on all parameters, in \autoref{tab:catalogs}. We also show corner plots for the highest weight models for in \autoref{fig:posteriors_powerlaw}, \autoref{fig:posteriors_truncated}, and \autoref{fig:posteriors_mdd}, for the \texttt{Powerlaw}, \texttt{Truncated}, and \texttt{MDD} catalogs, respectively. Examining the plots and the table, we can draw a number of conclusions. First, the method does an extremely good job at distinguishing whether mass-independent or mass-dependent disruption is a better fit to the data. The Akaike weights are unambiguous. Second, the method recovers the input parameters extremely accurately. The recovered mass function slopes $\alpha_M$ are accurate to better than $0.1$ in all cases, and the powerlaw indices describing the age distribution ($\alpha_T$ for the mass-independent cases, $\gamma_{\rm mdd}$ for the mass-dependent ones) are recovered with similar accuracy. We also recover the locations of breaks in the mass or age distributions with accuracies of $0.1 - 0.3$ dex, with the sole exception of the \texttt{Powerlaw} case, where by construction our sample should not be able to constrain $M_{\rm break}$. Indeed, in this case we find a very broad posterior PDF that rules out a break mass below $\sim 10^{5.5}$ $M_\odot$, but otherwise leaves the value unconstrained. In summary, we find that our new method offers excellent performance for data sets of a sample size and at an error level typical of modern observations.

\subsection{Sensitivity of the Method to Errors}
\label{ssec:mismatch}

\begin{figure}
\centerline{
\includegraphics[width=\columnwidth]{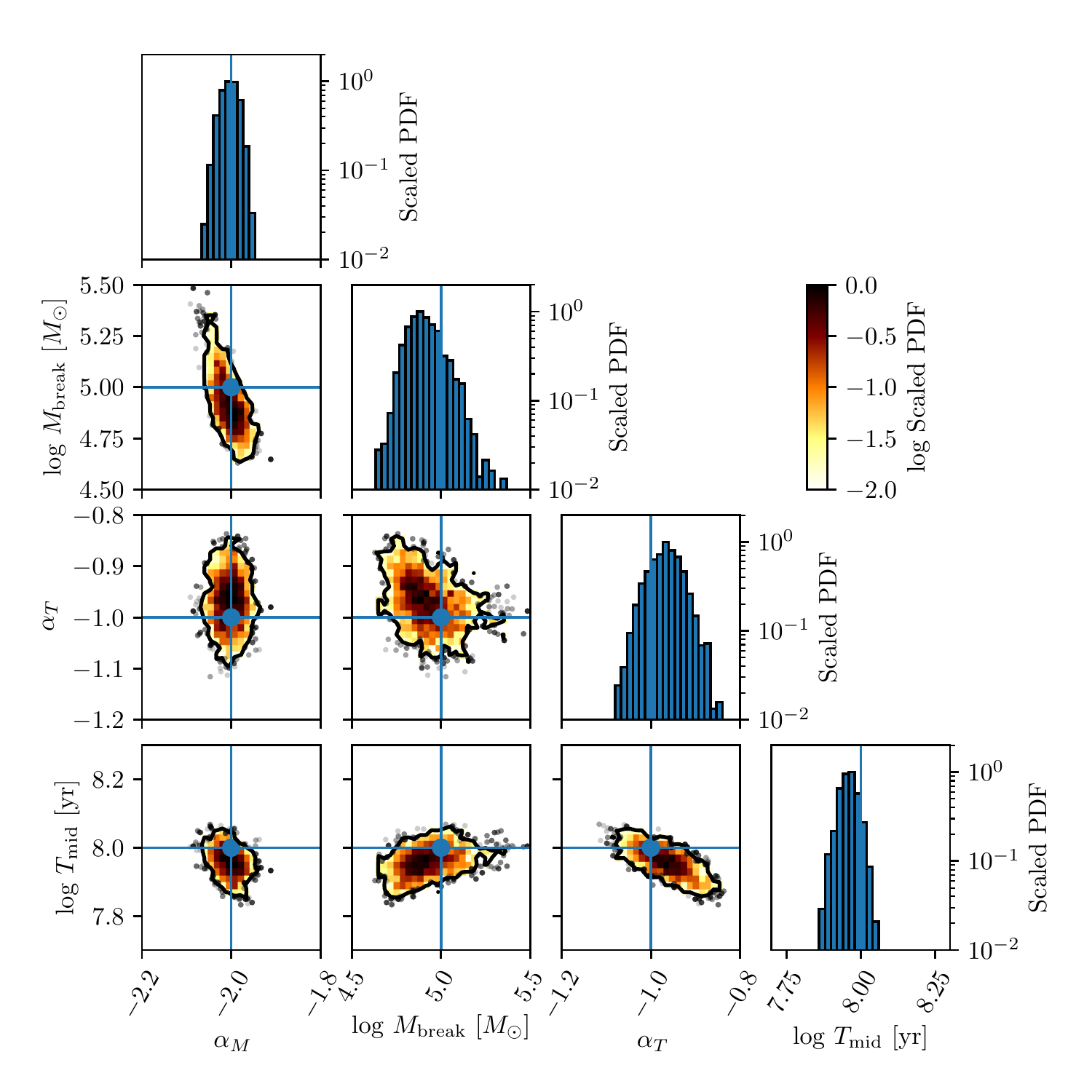}
}
\caption{
\label{fig:posteriors_err2}
Same as \autoref{fig:posteriors_truncated} for the \texttt{DoubleErr} catalog. Note that the ranges on the axes for this Figure are identical to those used in \autoref{fig:posteriors_truncated}, so the two may be compared directly.
}
\end{figure}

\begin{figure}
\centerline{
\includegraphics[width=\columnwidth]{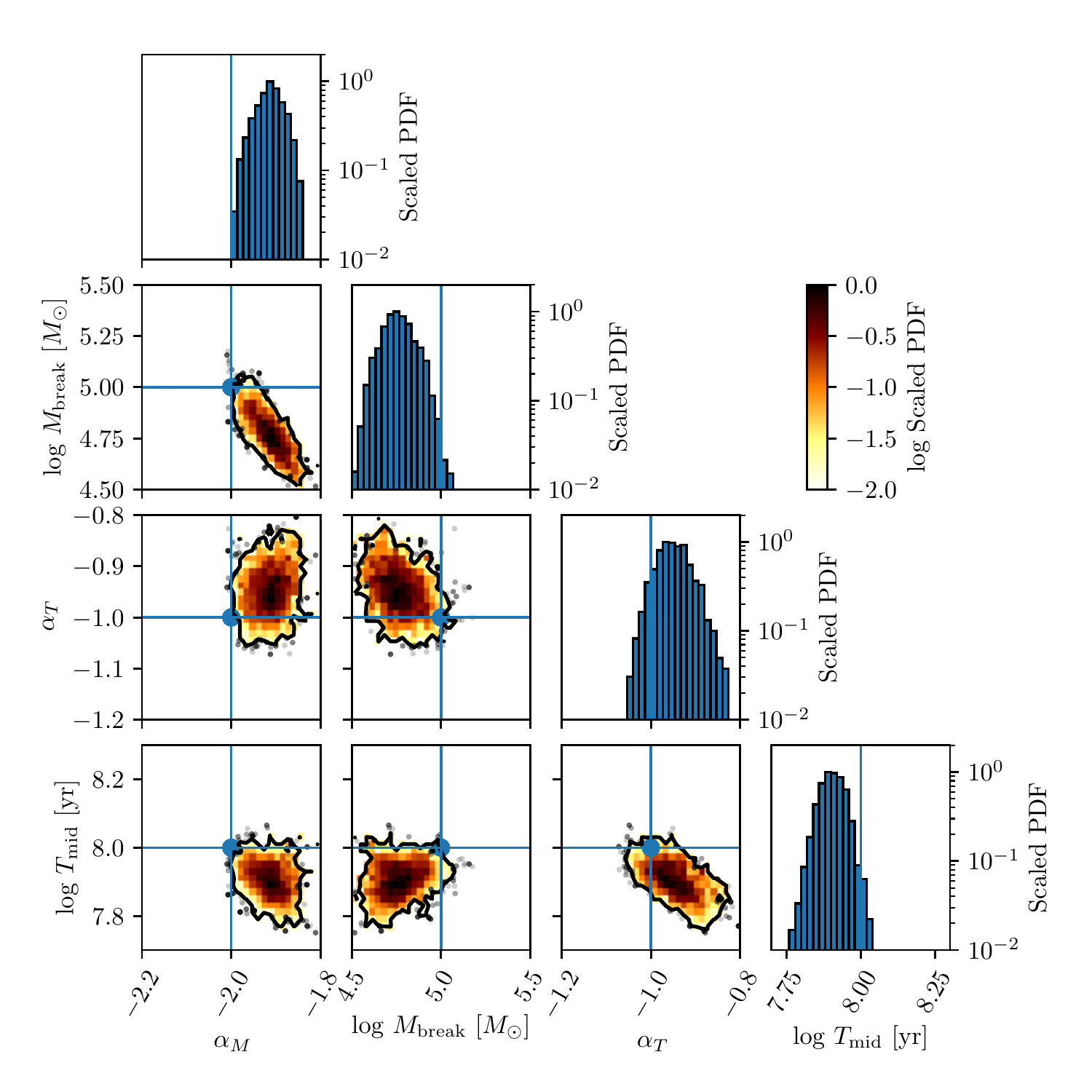}
}
\caption{
\label{fig:posteriors_mismatch}
Same as \autoref{fig:posteriors_truncated} for the \texttt{LibMismatch} catalog. Note that the ranges on the axes for this Figure are identical to those used in \autoref{fig:posteriors_truncated}, so the two may be compared directly.
}
\end{figure}

\begin{figure}
\centerline{
\includegraphics[width=\columnwidth]{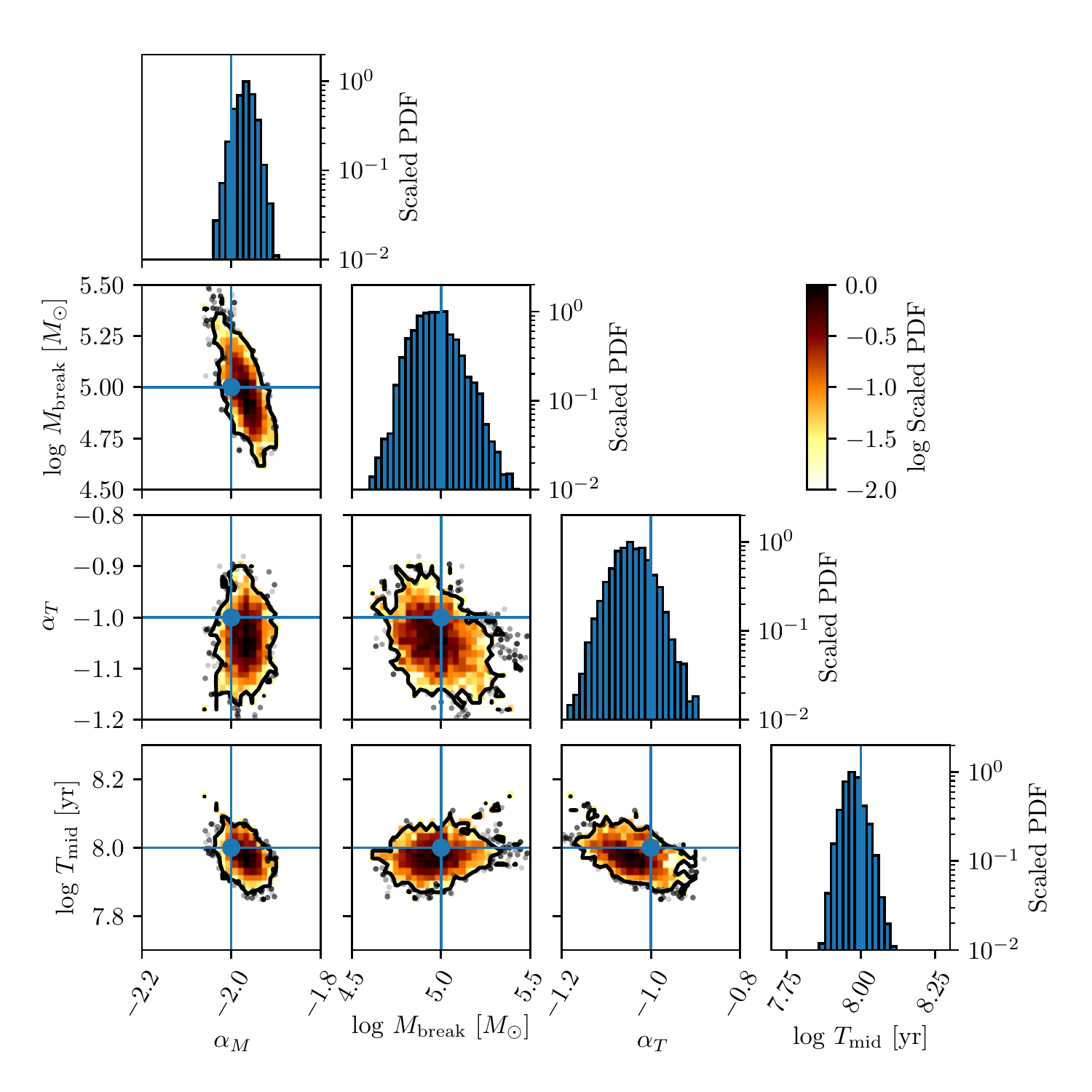}
}
\caption{
\label{fig:posteriors_comperr}
Same as \autoref{fig:posteriors_truncated} for the \texttt{CompMismatch} catalog. Note that the ranges on the axes for this Figure are identical to those used in \autoref{fig:posteriors_truncated}, so the two may be compared directly.
}
\end{figure}

We next investigate how robust our method is against various types of error, using the mock catalogs \texttt{DoubleErr}, \texttt{LibMismatch}, and \texttt{CompMismatch}. To remind the reader, each of these mock catalogs has the same physical parameters as \texttt{Truncated}, but differs in the errors in some way. \texttt{DoubleErr} has photometric errors of $0.2$ mag instead of $0.1$ mag, \texttt{LibMismatch} uses Padova rather than MIST tracks, as well as a starburst attenuation curve instead of a Milky Way extinction curve to model the effects of dust. Finally, \texttt{CompMismatch} uses a different completeness function than the one assumed in the analysis. We analyse each of these cases using the same procedure as described in \autoref{ssec:mock_analysis}; we run the MCMC in for 800 iterations instead of 500, and derive the posterior PDF from the final 300, since we find that it takes slightly longer for the posterior distributions to stabilise in at least some of these cases.

We report the marginalised posteriors in \autoref{tab:catalogs}, and show corner plots for the posteriors in \autoref{fig:posteriors_err2}, \autoref{fig:posteriors_mismatch}, and \autoref{fig:posteriors_comperr} for the \texttt{DoubleErr}, \texttt{LibMismatch}, and \texttt{CompMismatch} cases, respectively. As is clear from the Table and from comparing these three figures to \autoref{fig:posteriors_truncated}, in all three cases with increased errors we can still clearly distinguish between mass-independent and mass-dependent disruption, and still clearly identify the truncation in the mass function and the break in the age distribution. Doubling the photometric errors has remarkably little effect on the accuracy of the resulting fits, likely because, given the large number of clusters in the catalog, the limiting factor in the accuracy of the fits is stochastic sampling and the degeneracies it induces in the tracks, not the accuracy with which individual clusters' photometry can be measured. The main effect of using a library that does not precisely match the data is to induce a systematic shift in the posteriors, while leaving the shape and width of the posterior distribution largely unchanged. The best-fitting mass function and age distribution slopes for \texttt{LibMismatch} are displaced from their true values ($\alpha_M = -2$, $\alpha_T = -1$) to $\alpha_M=-1.9$, $\alpha_T = -0.95$, while the truncation mass and break in the age distribution are shifted to $\log \left(M_{\rm break}/M_\odot\right) = 4.76$ (true value $5.0$) and $\log\left(T_{\rm mid}/\mbox{yr}\right) = 7.9$ (true value 8.0). Using an incorrect completeness function (as in \texttt{CompMismatch}) has a similar effect, but smaller in magnitude. This may well be a function of our parameterisation: the effect of using an incorrect estimate of our completeness  is to induce artificial wiggles in the completeness-corrected luminosity function at low luminosity. However, since we are forcing the functional forms of our physical distributions to be powerlaws without any such wiggles, the wiggles do not greatly alter the best-fit powerlaw slopes. Overall our results suggest that fits to cluster demographics likely to be limited to an accuracy of $\approx \pm 0.1$ in the age and mass function slopes, and a few tenths of a dex in age or mass truncations.

It is important to note that the results of our fit to the \texttt{LibMismatch} catalog differ from both the true values and the estimates we find for the \texttt{Truncated} case by amounts that range from one to a few standard deviations. This implies that our method is sufficiently sensitive that, for our sample size of a few thousand clusters, the accuracy of the results is ultimately limited by the quality of the underlying physical models, rather than by the data quality or statistics. We emphasise that this is a highly non-trivial statement. Even for samples of this size, analysis using conventional methods generally yields results where the choice of tracks or libraries does not change the best fit values by more than the statistical error bars \citep[e.g.,][]{adamo17a}, and even using a method that fully accounts for stochastic sampling and returns the full posterior distribution, estimates of for the mass and age of individual clusters are relatively insensitive to the choice of underlying stellar model \citep{krumholz15c}. It is only because our method is capable of exploiting all the information present in a realistically-sized sample of clusters that we have reached a point where we are limited by systematic rather than statistical uncertainties.

\section{Comparison to Conventional Methods}
\label{sec:conventional}

We have now demonstrated that our new method is capable of recovering the parameters describing a cluster population with high accuracy, and that it is robust against plausible systematic and random errors. We now turn to the question of how well our method performs compared to more conventional approaches that do not involve full forward-modelling. We therefore re-analyse the \texttt{Powerlaw} and \texttt{Truncated} catalogs using a conventional method. There are a wide range of such methods, which differ from one another significantly in their details. We have chosen a methodology based on those used in a number of recent publications, and that should be broadly representative of the strengths and weaknesses of current techniques. As we will see below, the conventional method performs far worse when operating on the same data.

\subsection{Fitting individual clusters}

In any conventional method to derive cluster population properties, the first step is to derive the properties of the individual star clusters from their photometry assuming that the IMF is fully sampled, so that the relationship between photometry and physical properties is deterministic and we can assign a single best-fitting mass and age. To facilitate this we generate a grid of \slug~simulations of star clusters at a range of ages and extinctions with a fully-sampled IMF, i.e., with no stochasticity. We then find the best-fitting age and extinction for each catalog cluster by calculating the minimum $\chi^2$ between the colours of the model grid and the colours in the catalog, and find the best-fitting age by scaling the absolute magnitudes of the cluster to those of the model grid. We give a full explanation of the procedure in \autoref{app:chi2method}.

\begin{figure}
\includegraphics[width=\columnwidth]{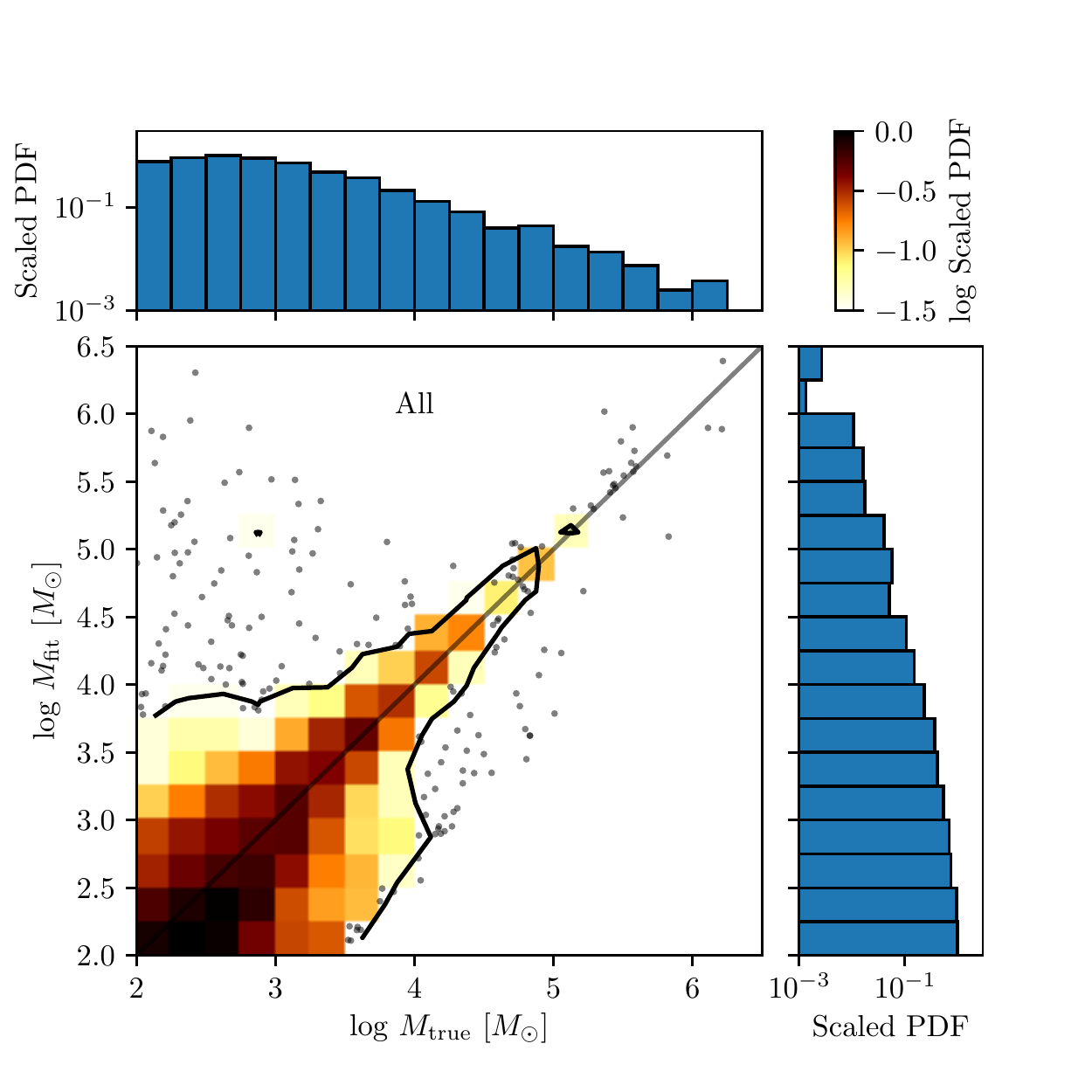}
\includegraphics[width=\columnwidth]{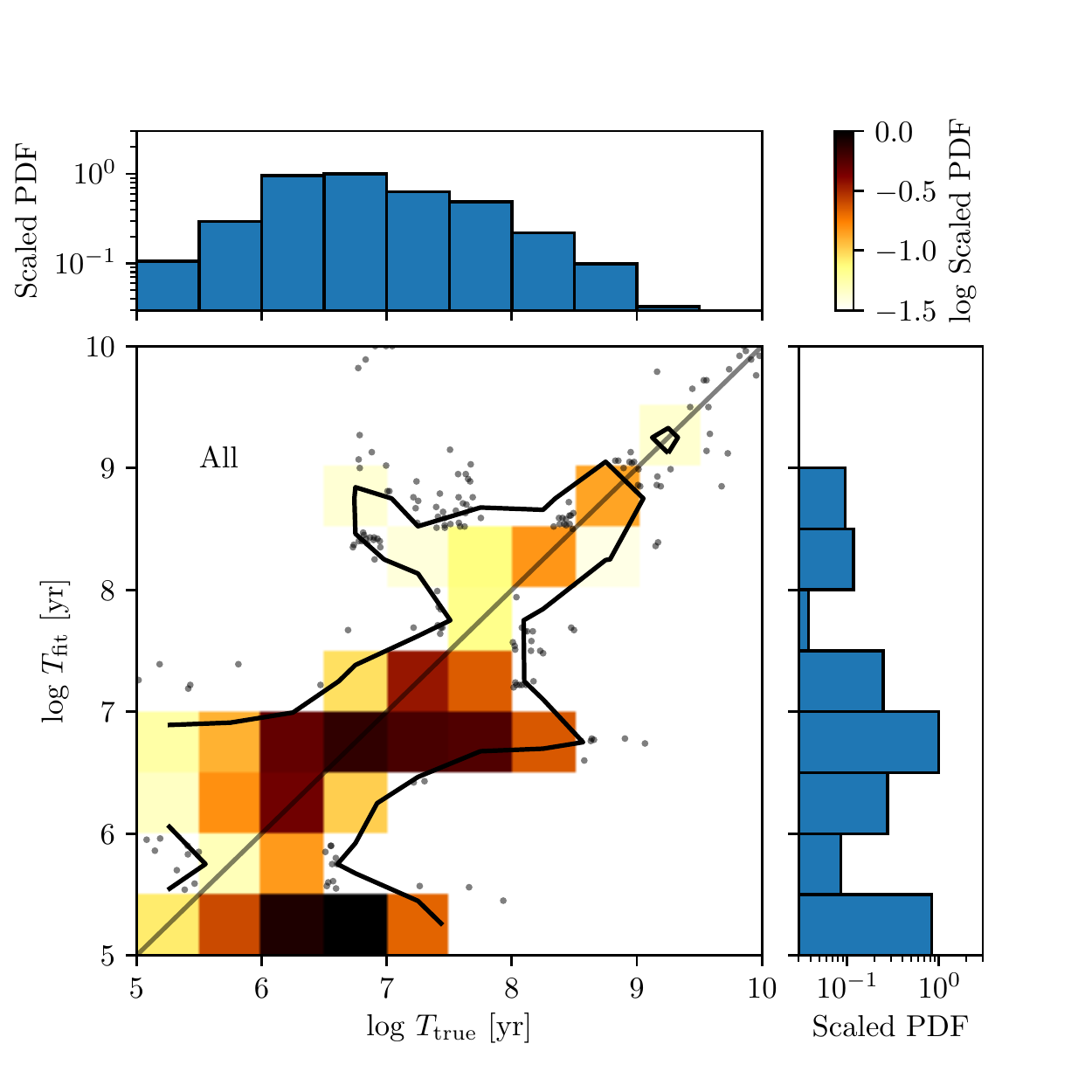}
\caption{
\label{fig:fit_all}
Comparison between true and best fit cluster masses and ages, for all clusters in the \texttt{Powerlaw} mock catalog for which the reduced $\chi^2$ value of the fit is $<5$. In the top panel the heat map shows the density of clusters, measured in bins 0.25 dex wide, in the plane of $\log M_{\rm true}$ versus $\log M_{\rm fit}$, where the former is the true cluster mass and the latter is the best fit determined from the procedure outlined in the main text. The black contour marks a density of $10^{-1.5}$ relative to the maximum, points show individual clusters in low-density regions, and the grey line shows the one-to-one line where a perfect fit would lie. The flanking histograms show the distributions of true and best fit mass. The bottom panel shows the analogous comparison between actual and best fit clusters ages, with bins 0.5 dex wide for the histograms.}
\end{figure}

To test how well this method performs, in \autoref{fig:fit_all} we compare the best fit and true masses and ages of the clusters in the \texttt{Powerlaw} mock catalog; results for other catalogs are qualitatively similar, and so for now we simply focus on the \texttt{Powerlaw} case. When making this figure, we only include clusters for which the reduced $\chi^2$ value ($=\chi^2/2$, since we have five photometric bands and three model parameters) of the fit is $<5$ on the grounds that larger values indicate a poor fit. This cut removes just under 20\% of the clusters in the mock catalog, and the results are not very sensitive to the exact threshold used to remove bad fits. We see that, in agreement with our expectations, the masses and ages recovered by the fitting procedure are for the most part accurate to within a factor of a few for clusters whose true mass is above $\sim 10^4$ $M_\odot$, or whose true age is above $10^8$ yr. In this mass and age range (which are largely overlapping in the observed catalog, since only a massive cluster will be bright enough to be detected at ages above $\sim 10^8$ yr), stochastic sampling of the IMF has minor effects, so full sampling is a reasonable approach. At lower masses and ages the scatter is substantially larger, but the majority of the data still cluster around the one-to-one line.

There are also a few extreme outliers very far from the one-to-one line. These typically have a true mass is below $\sim 1000$ $M_\odot$ and age $\sim 10^7$ yr, but the best fit mass is $>10^6$ $M_\odot$, and the best fit age above $\sim 10^9$ yr. This phenomenon occurs when stochastic sampling of the IMF in a low-mass, middle-aged cluster whose light is dominated by a small number of evolved stars happens to produce colours quite similar those of a much more massive, older stellar population whose light is dominated by a large number of low-mass stars. Because the conventional method does not include stochastic sampling, no models in the right age range match the colours as well as the older model, and thus the age and mass are catastrophically misestimated.

\subsection{Mass and age cuts}

To handle the problem of incompleteness, the next step in a conventional analysis is to cut the sample based on mass and age, restricting to ranges that are thought to be reasonably complete. We therefore next remove from the mock catalog any clusters whose best-fit age is $>10^{8.5}$ yr or whose best-fit mass is $<10^{3.75}$ $M_\odot$; these cuts are quite comparable to those normally used for star cluster analysis \citep[e.g.,][]{adamo17a, chandar17a, johnson17b}, and roughly corresponds to the region in age-mass space where almost all clusters will lie at $V \leq -5$ mag, and thus in the range where our catalog is complete. This combined with the cut on quality of fit reduces the \texttt{Powerlaw} catalog to 366 clusters (from the original 5,629), i.e., only $6.5\%$ of the original sample passes all the cuts. The reason that the mass-age cut so severely reduces the size of the available data is that, although any cluster with mass above $10^{3.75}$ $M_\odot$ and age below $10^{8.5}$ yr is very likely to have $V < -5$ mag and thus be bright enough to make it into the observed catalog, the converse is not true, i.e., having $V < -5$ mag in no way guarantees that the mass is above $10^{3.75}$ $M_\odot$ and age below $10^{8.5}$ yr. Indeed, because low-mass clusters are intrinsically much more common than more massive ones, the majority of the clusters bright enough to be observed have best-fitting masses that place them below our mass limit. They are bright enough to be observed because they are much younger than $10^{8.5}$ yr. We could retain more of these clusters by using a lower mass cut, but only at the price of having an even more severe age cut in order to ensure completeness. The tradeoff between age and mass cuts that we have made is comparable to the ones used by previous authors.

\begin{figure}
\includegraphics[width=\columnwidth]{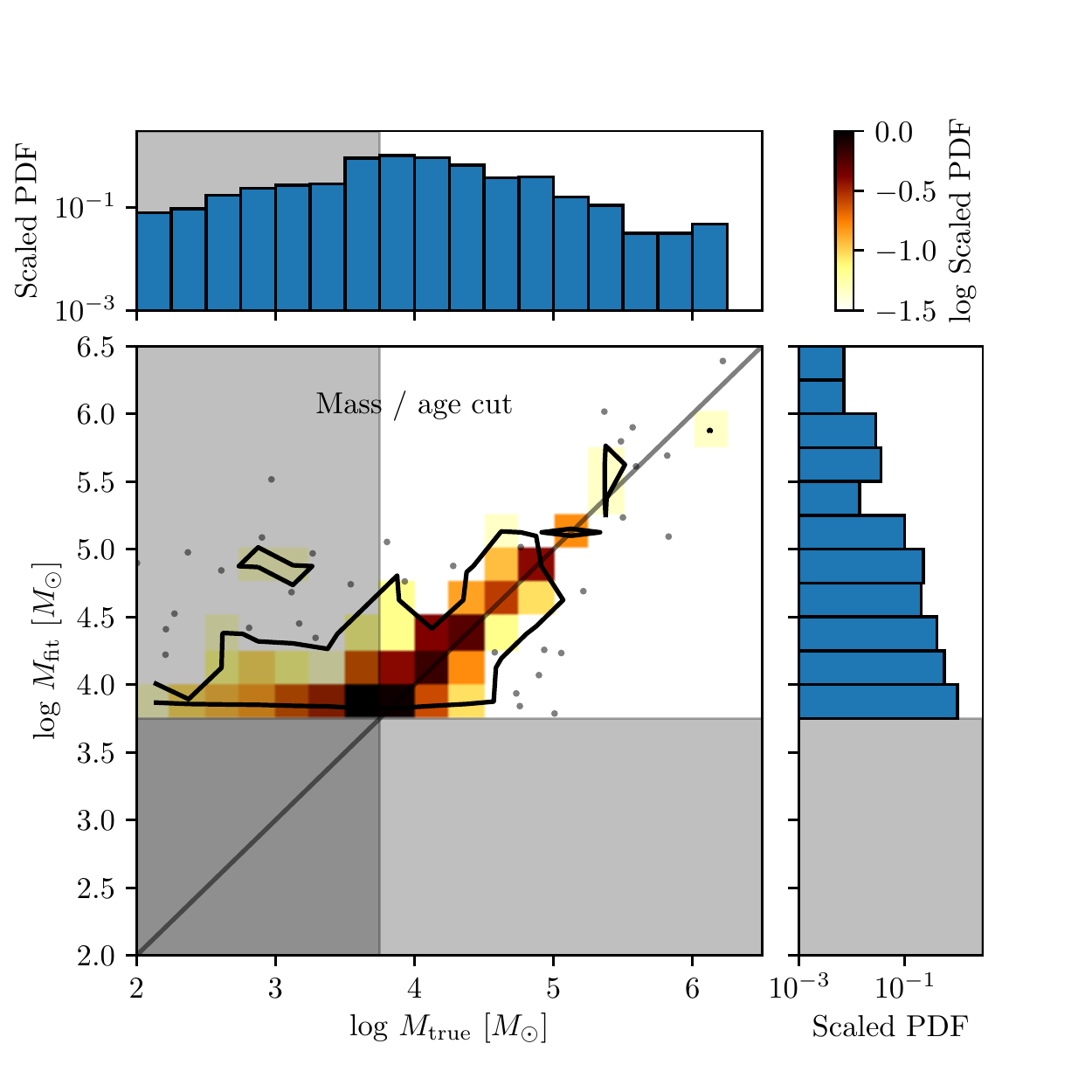}
\includegraphics[width=\columnwidth]{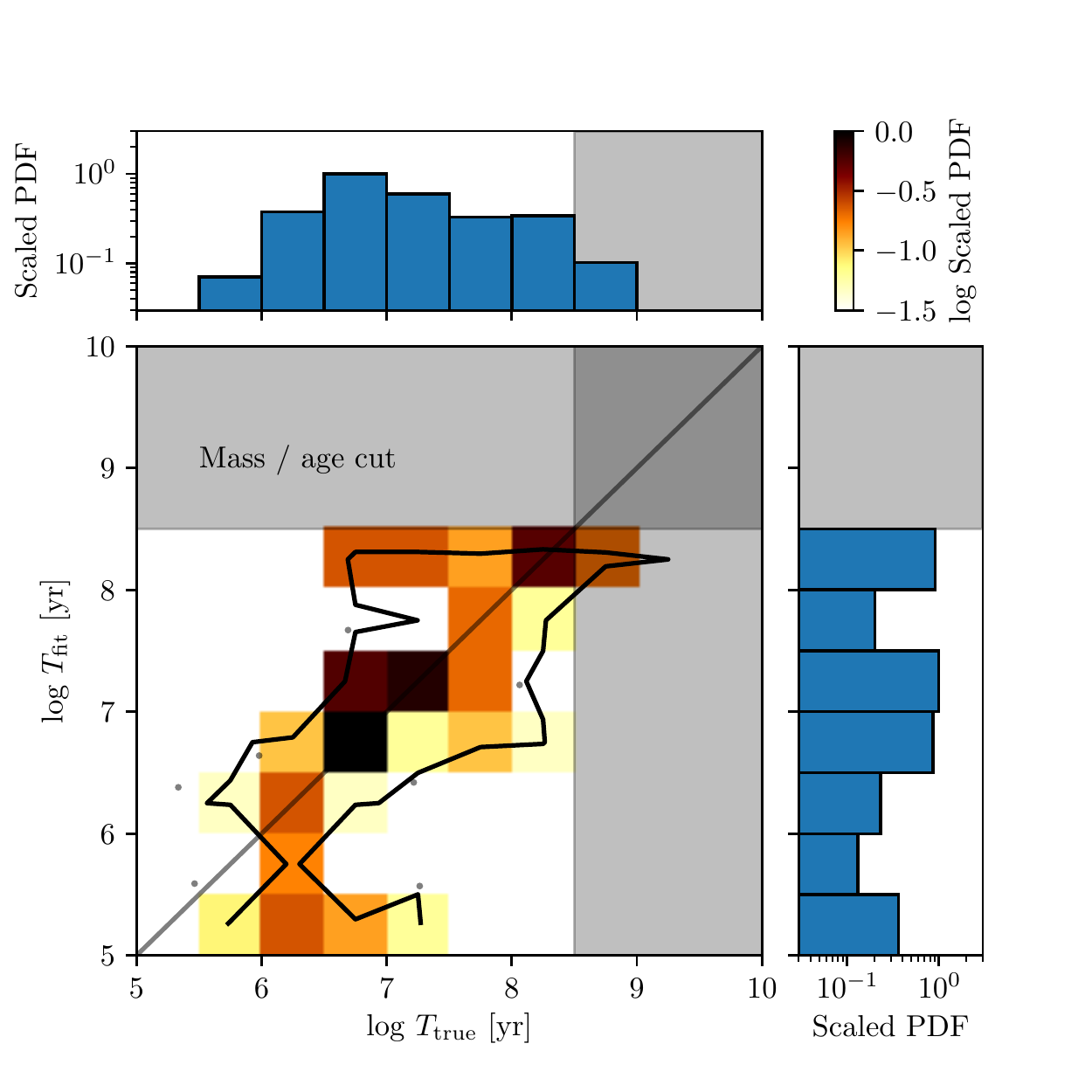}
\caption{
\label{fig:fit_cut}
Same as \autoref{fig:fit_all}, but now including only those clusters with a best-fit mass above $\log (M/M_\odot) = 10^{3.75}$ and best-fit age below $\log (T/\mbox{yr}) = 8.5$. The grey regions show the ranges of mass and age removed by the cut.
}
\end{figure}

We plot the distribution of mass and age for the remaining clusters for the \texttt{Powerlaw} catalog in \autoref{fig:fit_cut}. One particularly noticeable effect in this plot is that the lowest bins of fitted mass contain a very significant number of interlopers whose true mass is smaller, visible as the large tail of clusters extending into the grey region in the upper panel of \autoref{fig:fit_cut}. This is not a small effect: the true number of clusters in the mass range $\log M/M_\odot = 3.75 - 4.0$, only considering clusters that make it into the fitted sample (i.e., excluding those that have been dropped because their best-fit mass is too small, or because they do not have good $\chi^2$ values) is 64, while the fitting procedure produces 141, i.e., more than double the correct count. This bias has the effect of slightly flattening the best-fit mass distribution. Its origin lies in the fact that low-mass clusters are intrinsically more common than high-mass ones, and thus there are many more low-mass clusters scattering to high mass fits than high mass clusters scattering to lower masses.  The problem is worsened by the use of a mass cut, which creates an asymmetry: clusters whose fitted masses scatter below their true ones are preferentially removed from the sample, while those whose fitted masses scatter above their true ones are preferentially included.

\begin{figure}
\includegraphics[width=\columnwidth]{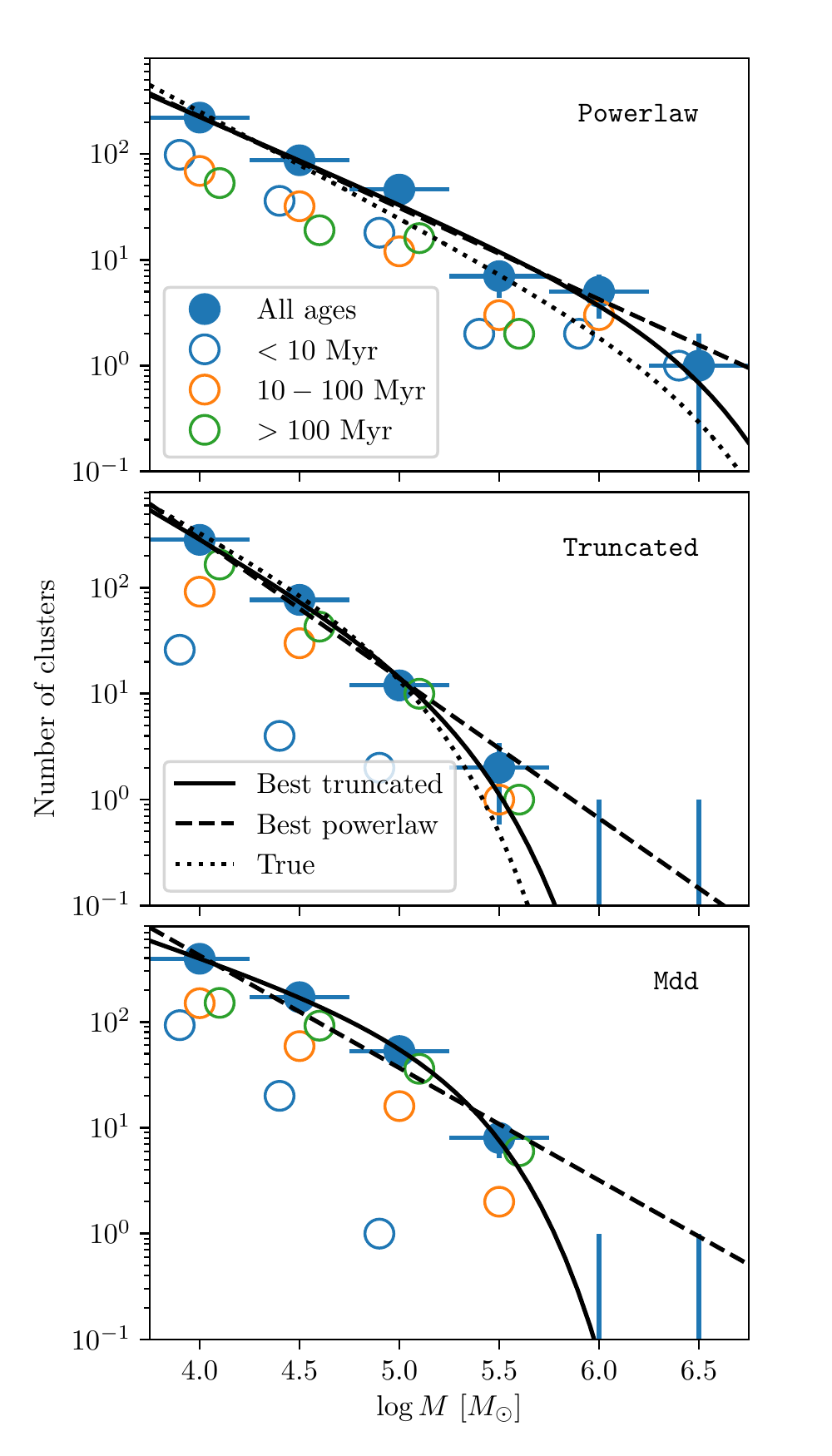}
\caption{
\label{fig:mfit_bins}
Binned mass distributions for the \texttt{Powerlaw}, \texttt{Truncated}, and \texttt{Mdd} catalogs (top to bottom, as indicated). Solid blue points indicate the number counts for clusters of all ages below $10^{8.5}$ yr; horizontal error bars show the width of each bin, and vertical error bars show Poisson errors on the number count. Open circles show the number counts for the same bins for clusters ages $<10$ Myr, $10-100$ Myr, and $>100$ Myr, as indicated; the horizontal positions of these points have been perturbed slightly, and error bars on them suppressed, to reduce confusion. The black solid and dashed lines show the best truncated- and pure-powerlaw fits to the solid points (see main text), while the black dotted line shows the true mass distribution, normalised to match the data in the lowest mass bin; we omit the true mass distribution for the \texttt{Mdd} case because it is not time-independent.
}
\end{figure}

\subsection{Fitting the distributions}

The final step in a conventional analysis is to fit the remaining data to constrain the properties of the population. Numerous fitting techniques exist in the literature, and it is not our goal here to perform a comprehensive comparison among them, so we select one representative example: we bin the data uniformly by mass and by age, using the binning shown in \autoref{fig:fit_cut}, then perform $\chi^2$ fits to the resulting binned distributions using the bin centres as the independent variable and bin number counts with Poisson errors as the dependent variable. A number of recent publications have used this method \citep[e.g.,][]{silva-villa14a, chandar15a, chandar17a}. For simplicity we only consider mass-independent disruption for this exercise, since for mass-dependent disruption we would need to fit the joint mass-age distribution, or fit multiple age distributions at constant mass or mass distributions at constant age.

For the mass distribution we fit to both exponentially-truncated and pure powerlaws (i.e., to functional forms $dN/dM \propto M^{\alpha_M}$ and to $dN/dM \propto M^{\alpha_M} e^{-M/M_{\rm break}}$). We plot the resulting best fits against the binned data and the true input distributions in \autoref{fig:mfit_bins}. We see that binning in mass does allow one to qualitatively recover roughly the correct mass distributions, but with significant defects compared to the forward-modelling method. First, the truncated-powerlaw fits return $\alpha_M = -1.82 \pm 0.11$, $\log M_{\rm break} = 6.47\pm 0.93$ ($1\sigma$ error bars) and $\alpha_M = -2.07\pm 0.10$, $\log M_{\rm break} = 5.22\pm 0.2$ for the \texttt{Powerlaw} and \texttt{Truncated} cases, respectively; compared to the results given in \autoref{tab:catalogs} the central values for $\alpha_M$ are significantly further from the true value ($\alpha_M = -2$), and the error bars are larger by a factor of $3-5$; one should also recall that even these error bars are underestimates, since they include only shot noise in the bin counts and not errors in the assigned masses.

Moreover, unlike the forward-modelling method, the binned method does not always successfully distinguish between pure- and truncated-powerlaw fits. For the \texttt{Powerlaw} catalog the reduced $\chi^2$ values for the truncated- and pure-powerlaw fits are $2.6$ and $2.1$, respectively, so the pure-powerlaw is a marginally better fit, as it should be since the data really do lack the statistical power to distinguish the two cases. However, for the \texttt{Truncated} catalog, the truncated- and pure-powerlaw fits have reduced $\chi^2$ values of $0.32$ and $0.95$ respectively, so the pure-powerlaw fit is actually preferred on the grounds that the truncated-powerlaw is overfitting the data. Indeed, one could have guessed this simply from \autoref{fig:mfit_bins}, since the pure-powerlaw fit falls within the Poisson error bars for every bin. For this catalog our new method successfully recovers an important result that the conventional one misses. The reasons for the new method's superior performance are obvious: the conventional $\chi^2$ method introduces a significant source of error by assigning each cluster a single best-fitting mass, discards more than 90\% of the sample due to the mass and age cuts it requires, then and discards even more information information by binning the data that remain.

\begin{figure}
\includegraphics[width=\columnwidth]{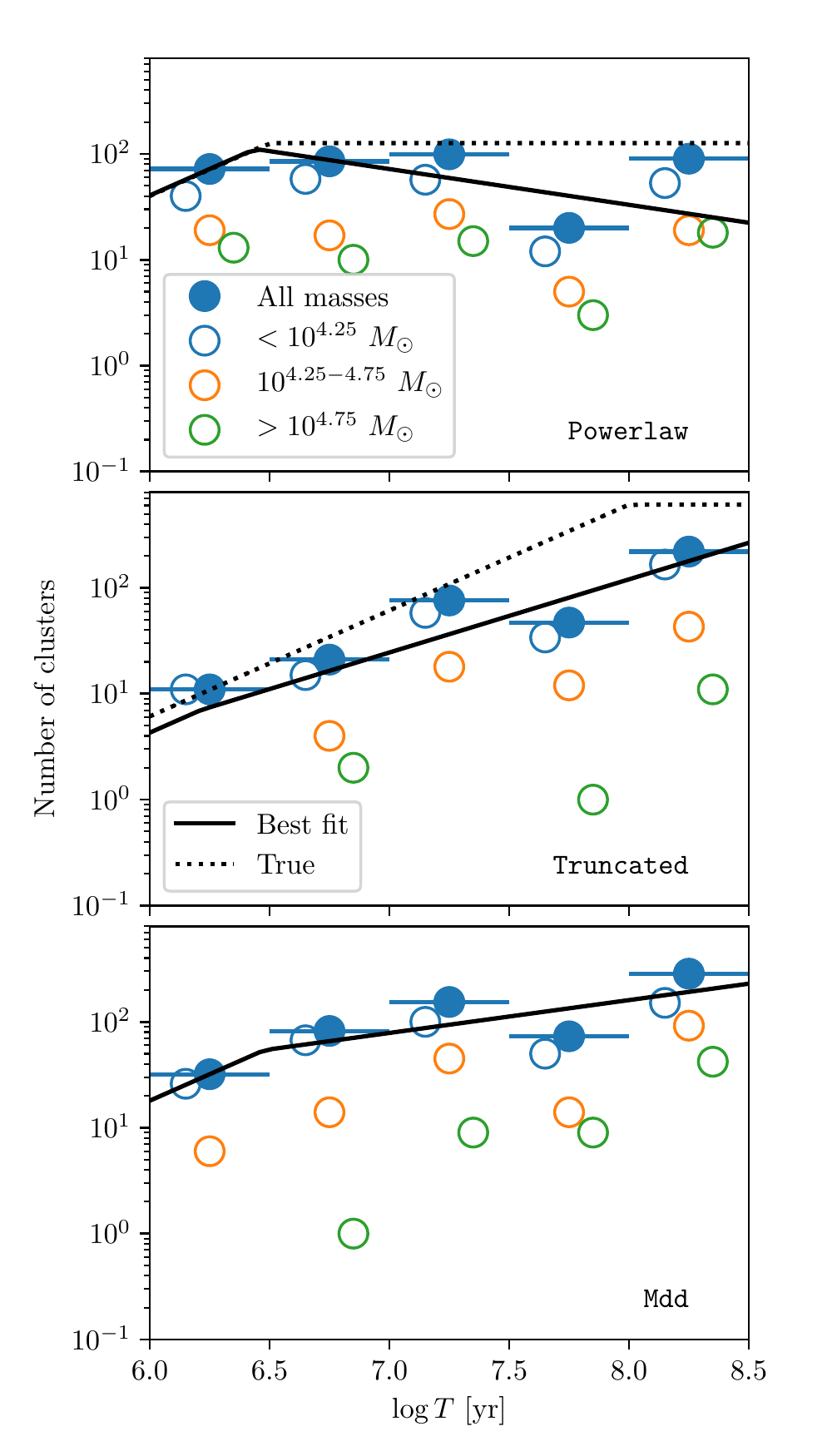}
\caption{
\label{fig:tfit_bins}
Same as \autoref{fig:mfit_bins}, but showing the binned age distributions instead of mass distributions.
}
\end{figure}

We show the analogous result for the binned age distributions in \autoref{fig:tfit_bins}; the fits shown are to a distribution $dN/dT \propto T$ for $T < T_{\rm mid}$ and $dN/dT \propto T^{\alpha_T}$ for $T>T_{\rm mid}$, with $T_{\rm mid}$ and $\alpha_T$ as our fit variables. Again, we see that the results of $\chi^2$ fitting have a rough qualitative resemblance to the input distributions, but that the properties recovered are considerably less accurate. For the \texttt{Powerlaw} catalog the best-fit values are $\alpha_T = -1.34 \pm 0.39$ and $\log T_{\rm mid} = 6.44 \pm 0.37$, while the true inputs are $\alpha_T = -1$, $\log T_{\rm mid} = 6.5$; the slope is poorly-determined in part due to the undercount in the $10^{7.5} - 10^8$ yr bin (also visible in \autoref{fig:fit_cut}), which drags the best fitting slope downward and results in a poor fit quality overall (reduced $\chi^2$ of 40). This is an artefact of trajectory of a fully sampled cluster through colour space in this age range, which makes it easy to mistake an older cluster for a younger one; such artefacts are known peril in $\chi^2$ fitting methods, as illustrated for example by the age-striping visible in Figure 10 of \citet{chandar10a} or Figure 14 of \citet{adamo17a}. Our new method avoids this problem by retaining the full posterior PDF rather than simply using the best fit, and \autoref{tab:catalogs} shows that our method recovers $\alpha_T$ with uncertainties an order of magnitude smaller: $\approx 0.02$ rather than $\approx 0.2$.

For the \texttt{Truncated} catalog, the conventional $\chi^2$ method returns a best-fitting slope of $\alpha_T = -0.31 \pm 0.29$ and is unable to determine a meaningful value of $\log T_{\rm mid}$ (formally the uncertainty in this parameter diverges), whereas the true inputs are $\alpha_T = -1.0$, $\log T_{\rm mid} = 8$. It is not surprising that the method fails to recover $\log T_{\rm mid}$, since the break in the age distribution in the input data lies very close to the age of $10^{8.5}$ yr above which we had to discard data due to incompleteness in the conventional method. The recovered slope $\alpha_T = -0.31 \pm 0.29$ is essentially within $1\sigma$ of the correct slope (which is 0) for the age range below $10^8$ yr covered by the remaining data. However, this only serves to illustrate a further advantage of our new method, that it does not require such severe cuts on the data to cope with incompleteness.

In summary, we find that the new method we have introduced outperforms the traditional one across all dimensions. It returns more accurate parameter estimates with much smaller uncertainties, and across a wider domain in cluster mass and age, and it successfully distinguishes between truncated and non-truncated mass distributions in cases where conventional methods cannot.

\section{Summary and conclusion}
\label{sec:conclusions}

We introduce a new forward-modelling method to determine the demographics of a population of star clusters from unresolved photometry. The basic idea of the method is to consider a proposed distribution of cluster masses and ages, apply kernel density estimation to a precomputed library of models weighted by the observational completeness to predict the observed luminosity distribution, and adjust the proposed mass and age distributions to optimise agreement between the observed and predicted luminosity distributions. Our method does not require that the data be binned, allows analysis of heterogenous data sets where not all regions have been observed to the same depth or with the same filters, does not require that data be limited to a particular range of mass or age, and naturally accounts for the effects of stochastic sampling of the stellar IMF. The method is computationally efficient enough that a catalog of star clusters comparable to those obtained via recent \textit{HST} campaigns can be analysed in approximately half a day of computing time on a workstation.

We test our method on synthetic data sets and show that we are able to recover correct demographics for the underlying populations with very high accuracy; typical statistical uncertainties on slopes of the mass and age distributions are only $\sim 0.01$ dex. Our method distinguishes between alternative models for star cluster age distributions, including truncated versus powerlaw mass functions, and mass-independent versus mass-dependent disruption, with very high confidence. The performance of our method compares very favourable with that of traditional $\chi^2$ fitting methods, whereby one obtains a best-fitting mass and age for each individual star cluster, and then fits the demographics of the population. We show that this method produces uncertainties on the slopes of mass and age distributions that are as much as an order of magnitude larger than our forward-modelling technique, and often lacks the statistical power to distinguish between alternative physical scenarios for star cluster demographics. By repeating our analysis with different set of stellar tracks and dust distributions, we find that systematic errors in the slopes of mass and age distributions are $\approx 0.1$ dex, significantly larger than the statistical uncertainties our method produces, and thus our method is as accurate as possible given our current knowledge of stellar evolution and interstellar dust physics. Indeed, it is not beyond the realm of possibility that application of this method with different libraries of models could be used to diagnose which stellar evolution models are best able to match reality.

We implement our method as part of the Stochastically Lighting Up Galaxies (\slug) suite of stochastic stellar population and statistics tools. The software for the method and for all the tests presented in this paper, along with the mock catalogs on which we performed our tests and the full outputs of our MCMC analysis, are available from the \slug~website at \url{http://www.slugsps.com/cluster-population-pipeline}.

In future work we will apply the method presented here to the large sample of star clusters produced by the LEGUS survey \citep{calzetti15a, adamo17a}. In addition to providing an analysis of star cluster demographics with considerably greater accuracy than any previous method, this will enable us to obtain much better estimates for the properties of individual clusters. \citet{krumholz15c} showed that the largest uncertainty in the posterior probability distributions for the masses and ages for individual star clusters is the prior distribution, i.e., one's starting estimate of the frequency with which particular masses and ages arise in the population. By combining the methods outlined in \citet{krumholz15c} with the population demographics we determine from the method implemented here, we will be able to mitigate this uncertainty considerably, thereby improving our estimates cluster by cluster as well for the population as a whole.

\section*{Acknowledgements}

MRK acknowledges support from the Australian Research Council's \textit{Discovery Projects} funding scheme (project DP160100695) and from NASA through grants from the Space Telescope Science Institute research (programs \#13256 and \#13364). This project was undertaken with the assistance of resources and services from the National Computational Infrastructure (NCI), which is supported by the Australian Government. MF acknowledges support by the Science and Technology Facilities Council, grant number ST/P000541/1. This project has received funding from the European Research Council (ERC) under the European Union's Horizon 2020 research and innovation programme (grant agreement No 757535). 

\bibliographystyle{mn2e}
\bibliography{refs}

\begin{appendix}

\section{Algorithm for Fast Evaluation of the Likelihood Function}
\label{app:algorithm}

Here we describe the algorithm we use for fast evaluation of the sum
\begin{equation}
\label{eq:likesum}
s = \sum_{j=1}^{N_{\rm lib}} w_j(\vectheta) \mathcal{N}(\vecL_i-\vecL_j \mid \vech')
\end{equation}
in the likelihood function (\autoref{eq:likelihood}). The algorithm proceeds in three steps. First, we arrange the $N_{\rm lib}$ clusters in the library in a KD-tree of $N_F$ dimensions based on their luminosities $\vecL_j$. (Recall that $\vecL_j$ has $N_F$ dimensions.) We maintain separate KD-trees for each set of filters that are present in the observed sample. For each node $k$ in the KD-tree we compute the bounding box, i.e., the smallest $N_F$-dimensional rectangular prism aligned with the cardinal axes that contains $\vecL_j$ for each cluster luminosity $\vecL_j$ in the node. We also record the weight $w_j(\vectheta)$ of each cluster, and the summed weight $w_k$ of all the clusters contained in each node. This is an order $N_{\rm lib} \ln N_{\rm lib}$ operation, but need only be done once at the start of the calculation.

The second step in the algorithm is that, when we wish to change model parameters $\vectheta$, we recalculate the weights $w_j(\vectheta)$ of each individual cluster, and the summed weights $w_k$ of all nodes in the tree. This is an order $N_{\rm lib}$ operation, but need only be done once for each set of trial parameters $\vectheta$, not once per cluster, and can be parallelised trivially. On a workstation-level machine, using library of $10^7$ clusters, this step requires a few tenths of a second.

The third step is evaluation of the sum given by \autoref{eq:likesum} for each observed cluster luminosity $\vecL_i$. We carry out this step via a divide and conquer algorithm of order $\ln N_{\rm lib}$:
\begin{enumerate}
\item Let \texttt{nodes} be a list of nodes in the tree; for each node $k$ in the list, we record an estimate $s_k$ of its contribution to the sum and an upper bound $\Delta s_k$ on the error in this estimate, computed whenever a node is added to the list in step (iii). That is, for any node $k$, the contribution of the clusters within that node to $s$ is strictly bounded between $s_k - \Delta s_k$ and $s_k + \Delta s_k$.
\item Evaluate the current estimate of the sum $s = \sum_k s_k$ and the upper bound on the error $\Delta s = \sum_k \Delta s_k$, where the sums run over all nodes in the list \texttt{nodes}. If $\Delta s/s$ is smaller than some specified tolerance, stop iterating and return $s$. If not identify the node $k$ with the largest value of $\Delta s_k$.
\item Remove the node with the largest $\Delta s_k$ from \texttt{nodes}, and add its left and right children, which we denote $\ell$ and $r$, to \texttt{nodes}. Compute $s_{\ell}$ and $\Delta s_\ell$ as follows (and similarly for $s_r$ and $\Delta s_r$):
\begin{itemize}
\item If node $\ell$ is a leaf (i.e., it has no children), set $s_\ell = \sum w(\vectheta) \mathcal{N}(\vecL_i - \vecL_j \mid \vech')$, where the sum runs over all clusters in the leaf. Set $\Delta s_\ell = 0$. That is, if the node is a leaf, directly evaluate the contribution to the sum of all clusters in that leaf, and set the maximum possible error to zero.
\item If node $\ell$ is not a leaf, find the vectors $\Delta\vecL_{\rm near}$ and $\Delta\vecL_{\rm far}$ between $\vecL_i$ and the nearest and farthest points in the bounding box of node $\ell$, where distances are measured in units of $\vech'$. That is, $\Delta \vecL_{\rm near} = \vecL_i - \vecL_b$ for the point $\vecL_b$ within the bounding box of node $\ell$ that minimises $d = \sum_{n=1}^{N_F} (L_{i,n} - L_{b,n})^2 / {h'_n}^2$; similarly, $\Delta\vecL_{\rm far}$ is computed for the point $\vecL_b$ that maximises $d$. Note that if $\vecL_i$ is inside the bounding box, then $\Delta \vecL_{\rm near} = \mathbf{0}$. Set
\begin{eqnarray}
s_\ell & = & w_\ell \frac{\mathcal{N}(\Delta\vecL_{\rm near} \mid \vech') + \mathcal{N}(\Delta\vecL_{\rm far} \mid \vech')}{2} \\
\Delta s_\ell & = & w_\ell \frac{\mathcal{N}(\Delta\vecL_{\rm near} \mid \vech') - \mathcal{N}(\Delta\vecL_{\rm far} \mid \vech')}{2}.
\end{eqnarray}
Intuitively, this amounts to finding the maximum (minimum) possible contribution of the clusters in node $\ell$ to the sum, which would occur if all the clusters in that node were at the nearest (farthest) point within the bounding box. We then set the central estimate of the contribution of that node to the sum to the average of the minimum and maximum possible contributions, and the error to half the distance between them.
\end{itemize}
\item Go back to step (ii).
\end{enumerate}
The algorithm begins by adding the root node of the KD-tree to the list \texttt{nodes}, with $s_k$ and $\Delta s_k$ for it evaluated as in step (iii). We then iterate until the convergence condition in step (ii) is satisfied. The algorithm is efficient because (1) it quickly eliminates parts of the KD-tree that are far from $\vecL_i$, and thus make small contributions to $s$ and $\Delta s$, and (2) it removes the need to examine individual clusters whose separation in luminosity space is $\ll \vech'$, since these will be grouped into the same node, and nodes whose bounding boxes are $\ll \vech'$ in size will have $\Delta s_k \ll s_k$. In practice, we find that evaluation of $s$ to 1\% accuracy requires examining hundreds to thousands of nodes, depending on the number of filters $N_{\rm F}$, the bandwidth $\vech'$, and the number of points $N_{\rm lib}$ in the library, and the density of library points in the vicinity of $\vecL_i$. In our tests with 5 filters, a bandwidth of $0.05 - 0.1$ mag, and $10^7$ library points, typical evaluation times were $\sim 100$ $\mu$s per cluster on a workstation.

\section{Method for $\chi^2$ Fitting}
\label{app:chi2method}

Here we explain in detail the method we use to fit the mass, age, and extinction for each catalog cluster for our conventional analysis. The first step is to generate a grid of \texttt{slug} models with a fully sampled IMF. We use a fixed cluster mass of $10^6$ $M_\odot$, since in the non-stochastic case the photometry can be rescaled trivially to any chosen mass; specifically, if our \slug~run with a mass of $10^6$ $M_\odot$ produces a magnitude $m_{i,\rm model}$ in filter $i$, then the corresponding prediction for a cluster of mass $M$ is simply $m_i = m_{i,\rm model} - 2.5 (\log M - 6)$. We output model predictions $m_{i,\rm model}$ at a set of times from $10^5 - 10^{10}$ yr, with outputs spaced at $0.01$ dex. We repeat this calculation at dust extinctions from $A_V = 0 - 3$ mag in steps of 0.05 mag. The result is a grid of 30,561 models, each with predicted photometry in the same five bands as our mock catalogs as a function of cluster age and extinction. 

To fit obtain the best-fitting mass, age, and extinction, we proceed as follows. First, for each catalog cluster we find a best-fitting mass at each point in our age-extinction grid by finding the value $M_{\rm fit}$ that minimises
\begin{equation}
\chi^2 = \sum_i \frac{\left[m_{i,\rm cat} - m_{i,\rm model} + 2.5(\log M_{\rm fit} - 6)\right]^2}{\sigma_i^2},
\end{equation}
where $m_{i,\rm cat}$ is the magnitude of the catalog star cluster in band $i$, $\sigma_i$ is the error on this value ($0.1$ mag for all our mock catalogs), and $m_{i,\rm model}$ is the magnitude for the model grid point. Note that the required value of $M_{\rm fit}$ can be obtained analytically simply by solving the equation $d\chi^2/d\log M_{\rm fit} = 0$. We record the corresponding minimum value of $\chi^2$ at each grid point for each cluster. Second, for each cluster  in the mock catalog, we then find the grid point that produces the smallest $\chi^2$ value, which we denote $\chi^2_{\rm min}$. We assign the age and extinction of that point as the best fitting values for that cluster, and the corresponding mass recorded for that grid point as its best-fitting mass. Third, to derive the 68\% confidence interval, we find all the grid points for which $\chi^2 < \chi^2_{\rm min} + 2.3$, where the factor 2.3 comes from the numerical experiments of \citet{avni76a}, and find the minimum and maximum mass, age, and extinction among them.

Note that our method for deriving cluster mass and age identical to that used in \citet{adamo10a}, which has been used for star cluster analysis by a number of authors \citep[e.g.,][]{silva-villa14a, adamo15a, adamo17a}. We have developed our own code rather than using the \citet{adamo10a} code because that code uses simple stellar populations computed with \texttt{Yggdrasil} \citep{zackrisson11a}, whereas we wish to use photometry generated by \slug~so that our treatment of stellar evolution, stellar atmospheres, nebular emission, and dust extinction is identical to that used to generate the mock data and the libraries used for our new method. However, we have verified that the differences in the cluster masses and ages derived by our code versus the \citet{adamo10a} code are for the most part within the error bars.

\end{appendix}

\end{document}